\documentclass[aps,pre,twocolumn,notitlepage,10pt,longbibliography]{revtex4-2}

\usepackage{ulem}
\usepackage{amsmath,amssymb,amsfonts}
\usepackage{graphicx}
\usepackage{xcolor}
\usepackage{array}
\usepackage{booktabs,tabularx}
\usepackage{microtype}
\usepackage{bm}
\setcitestyle{numbers,sort&compress}

\usepackage[unicode,colorlinks=true]{hyperref}
\usepackage[nameinlink,noabbrev]{cleveref}

\newcommand{\Pvec}{\boldsymbol{\mathcal{P}}}

\begin{document}

\title{Intrinsic stochasticity in cell polarity and contact inhibition of locomotion}

\author{Mariia Kryvoruchko$^{1}$}
\author{Brian A. Camley$^{1,2}$}

\affiliation{$^1$Thomas C. Jenkins Department of Biophysics, Johns Hopkins University, Baltimore, MD 21218, USA}
\affiliation{$^2$William H. Miller III Department of Physics \& Astronomy, Johns Hopkins University, Baltimore, MD 21218, USA}

\begin{abstract}
When cells collide, they often exhibit ``contact inhibition of locomotion'' (CIL), a behavior in which cells repolarize and migrate away from the site of contact. Experimental CIL outcomes are highly variable --- why? Here, we develop a minimal stochastic model to quantify how intrinsic noise in cell polarity,  arising from the finite number of signaling molecules,  influences CIL decision-making. 
We simulate polarization dynamics by tracking individual Rho GTPase proteins that diffuse and switch stochastically between the cell membrane and cytosol. In the absence of cell--cell contact, the polarity axis diffuses rotationally --- the cell's orientation wanders ---  with a diffusion coefficient that decreases as Rho GTPase copy number increases.
Assuming that cell--cell contact inhibits Rho GTPase activation, we investigate how contact geometry, duration, and strength affect CIL sensitivity.
At low protein copy number, weak, brief, or spatially narrow contacts are masked by molecular noise. In contrast, at high protein copy number, intrinsic polarity noise is negligible, and randomness in CIL response is more likely to reflect the variability from collision to collision in the cell--cell contact properties.
\end{abstract}

\maketitle
\section{Introduction}

Contact inhibition of locomotion (CIL) is a fundamental mechanism in which migrating cells change direction upon contact with another cell. CIL contributes to wound healing~\cite{Abercrombie1954}, neural crest migration~\cite{Theveneau2011}, chemotaxis of groups of cells~\cite{Stramer2016, Theveneau2010}, and cancer metastasis~\cite{Abercrombie1976, Astin2010}. CIL arises from local interactions between individual cells --- contact leads to spatially localized signaling that suppresses protrusion at the contact site~\cite{CarmonaFontaine2008, Roycroft2015}. This signaling can be initiated via Ephrin/Eph or cadherin binding~\cite{Poliakov2004, Scarpa2016}, and sets off a cascade that ultimately governs whether a cell retracts, repolarizes, or continues migrating~\cite{Theveneau2010}. Experimental assays find that isolated motile cells often respond to contact with neighbors by changing direction~\cite{Mayor2010} --- but these responses are probabilistic. In 1D micropatterned environments, cells exhibit a limited set of outcomes: adhesion, repulsion (repolarization), train formation, or walk-past motion~\cite{Scarpa2013, Singh2021, Davis2015}. In contrast, in 2D, cells can repolarize across a continuous range of angles~\cite{Scarpa2016}. %
Our question is: \textit{what causes the noise in CIL responses}? Noise in biological contexts often comes from finite molecule number (``intrinsic'' noise), cell-to-cell heterogeneity, or variations in external perturbations ~\cite{Ladbury2012}. For CIL specifically, Wang et al.~\cite{Wang2024} showed that the observed  variability likely does not arise from intrinsic receptor--ligand binding noise. This suggests that stochasticity may instead originate downstream of the initial cell--cell contact, e.g. in cell polarity. Here, we will explore whether finite-copy-number fluctuations in cell polarity can drive noise in CIL response. 

Previous CIL modeling has included reaction-diffusion (RD) models~\cite{Camley2014,Kulawiak2016,Merchant2018,luo2025cell}, agent-based simulations ~\cite{Schnyder2017,Smeets2016,Camley2016,Zimmermann2016,Davis2012}, mechanical~\cite{Ron2023} and mechanochemical models~\cite{Levandosky2024}. 
However, these approaches typically treat noise phenomenologically, introducing randomness in a coarse-grained or abstract manner without linking it to underlying molecular processes.  As a result, they cannot capture how variability in protein copy number impacts cell decision-making.

Here, we extend  the Rho GTPase models of~\cite{Mori2008,Mori2011,Walther2012} to develop a minimal 1D stochastic model of cell polarity and CIL. Rho GTPases are key signaling molecules that regulate cytoskeletal dynamics and cell polarity, thereby controlling cell motility~\cite{EtienneManneville2002}. In our model, individual Rho GTPase molecules are tracked explicitly as they diffuse through the cell and switch between membrane-bound and cytosolic states. 
We study how fluctuations arising from finite molecule numbers affect the persistence of cellular polarization and a cell's sensitivity to contact. We find that at high copy number, polarization is stable and {contact response is robust.} 
On the order of a few thousand molecules, polarity becomes unstable, and cells are unable to reliably distinguish external signals from intrinsic fluctuations. These results identify conditions under which molecular noise masks contact-induced signals.
For fixed intrinsic noise, contact sensitivity decreases when contact occurs far from the leading edge of the cell, is short in duration, or weak in strength. Nevertheless, our simulations show that cells can still reliably detect contact  even if the contact is just a few microns wide, even at molecule numbers below experimentally measured values. 

\section{Methods and Materials}
\addcontentsline{toc}{section}{Methods and Materials}
\label{sec:methods}

\begin{figure}[ht]
  \centering
  \includegraphics[width=1\linewidth]{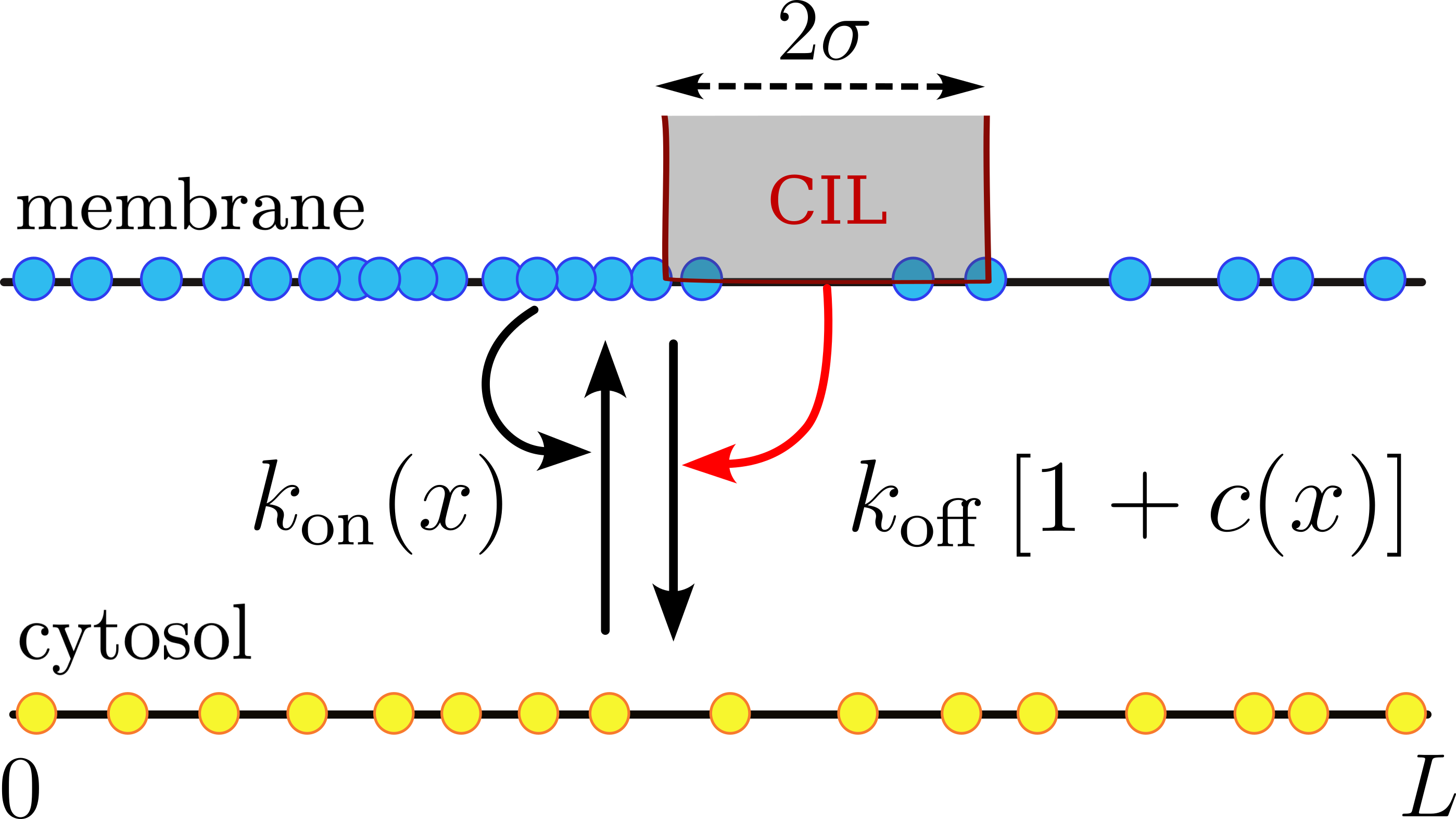}
  \caption{
  Schematic of the stochastic simulation illustrating protein switching between cytosolic and membrane-bound state. Binding occurs with rate $k_{\text{on}}$, which is regulated by positive feedback from the membrane-bound state,  while unbinding occurs with rate $k_{\text{off}}$, which can be locally enhanced by $c(x)$ (see Eq.~\ref{eq:koff(x, t)}) to represent cell--cell contact.
  }
  \label{fig:simulation_scheme}
\end{figure}

We develop a stochastic version of the classic wave-pinning reaction-diffusion model of Rho GTPase dynamics ~\cite{Mori2008,Mori2011} and extend it to account for cell--cell contact-induced repolarization. Rather than treating protein concentrations as deterministic continuous fields, as in ~\cite{Mori2008}, we explicitly model individual molecules and simulate their diffusion together with stochastic binding and unbinding dynamics. Our approach follows that of Walther \textit{et al.}~\cite{Walther2012}, but introduces {cell--cell contact (Fig.~\ref{fig:simulation_scheme})} and a distinct treatment of scaling total Rho GTPase number.

\begin{figure}[ht]
  \centering
  \includegraphics[width=1\linewidth]{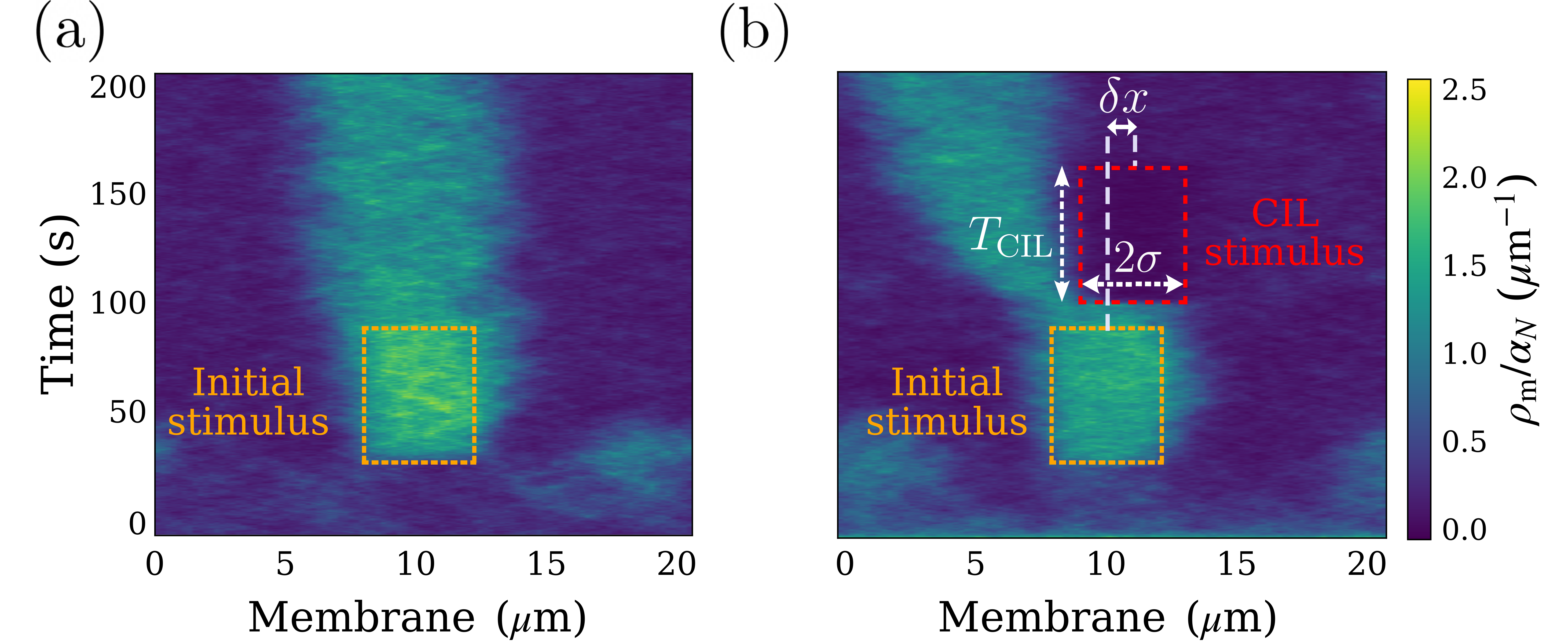}
  \caption{
 Kymographs of the normalized membrane concentration 
  $\rho_{\mathrm{m}}(t,x)/\alpha_{N}$,  with green intensity indicating a higher abundance of membrane-bound proteins: 
  \textbf{(a)} a polarized state in a freely migrating cell and 
  \textbf{(b)} repolarization in response to cell--cell contact, shown for
  $T_{\mathrm{CIL}} = 60\,\mathrm{s}$, 
  $\delta x = 1.5\,\mu\mathrm{m}$, 
  $\sigma = 1.5\,\mu\mathrm{m}$, 
  and $c_0 = 1$. 
The initial polarity before the CIL event is established by an initial stimulus
  (see Sec.~\ref{Stochastic model of Rho GTPase dynamics}; yellow dashed lines).
  }
  \label{fig:free_vs_CIL}
\end{figure}

\subsection{Stochastic model of Rho GTPase dynamics}
\label{Stochastic model of Rho GTPase dynamics}

We discretize time using a fixed time step \(\Delta t \) (see Appendix~\ref{appendix:varied_dt}) and initialize the system at \(t = 0~\mathrm{s}\) with approximately \(15\,\%\) of the total proteins on the membrane~\cite{Mori2008, Mori2011} and the remaining \(85\,\%\) in the cytosol. %
At each time step, every protein undergoes two stochastic processes: \textbf{(i)} diffusion with diffusion coefficients \(D_{\mathrm{m}}\) on the membrane and \(D_{\mathrm{c}}\) in the cytosol, and \textbf{(ii)}  exchange between the cytosolic (inactive) and membrane-bound (active) states with binding and unbinding rates \(k_{\mathrm{on}}\) and \(k_{\mathrm{off}}\), respectively (Fig.~\ref{fig:simulation_scheme}). Together, these processes determine the local membrane-bound protein density \(\rho_{\mathrm{m}}(x)\), which in turn modulates feedback through \(k_{\mathrm{on}}(x)\).

 We simulate a total of \(N\) proteins uniformly distributed along a membrane of length \(L\), {with protein position $x_k$  (\(k=1,\dots,N\)) on the periodic domain}. Each protein $k$  undergoes Brownian motion:
\begin{equation}
    x_k(t+\Delta t) \;=\; x_k(t) + \sqrt{2D\,\Delta t}\,\xi_k, 
    \qquad \xi_k \sim \mathcal{N}(0,1),
    \label{eq:Brownian_motion}
\end{equation}
where \(D\) is chosen as \(D_{\mathrm{m}}\) or  \(D_{\mathrm{c }}\) depending on whether the protein is membrane-bound or cytosolic. 
While many implementations of wave-pinning  approximate the cytosol as well-mixed when \(D_{\mathrm{c}} \gg D_{\mathrm{m}}\) ~\cite{Camley2017,Singh2022}, we explicitly simulate cytosolic diffusion on the 1D domain. %

Transitions between cytosolic and membrane-bound states are implemented as stochastic events  with position-dependent binding probability $p_{\mathrm{on}}(x)$ and a baseline unbinding probability $p_{\mathrm{off}}$ in the absence of contact
\begin{equation}
p_{\mathrm{on}}(x) = k_{\mathrm{on}}(x) \Delta t, 
\qquad 
p_{\mathrm{off}} = k_{\mathrm{off}} \, \Delta t.
\end{equation}
The binding rate $k_{\mathrm{on}}(x)$ incorporates positive feedback with a Hill form~\cite{Mori2008,Jilkine2011}:
\begin{equation}
    k_{\mathrm{on}}(x) = k_0 + k_1\,\frac{\rho_{\mathrm{m}}(x)^2}{\rho_{\mathrm{m}}(x)^2 + {K_N^{2}}}, 
    \label{eq:kon}
\end{equation}
where \(k_0\) is the basal binding rate, \(k_1\) the maximal feedback strength, and \(K_N\), which we discuss more below, sets the scale of the saturation of positive feedback (Table~\ref{Table:parameters}). Here, $\rho_{\mathrm{m}}(x)$ is the local membrane-bound protein density (Sec.~\ref{subsec: Domain Discretization and System Rescaling} shows how this is computed). After each time step, all diffusion and reaction updates are completed, \(\rho_{\mathrm{m}}(x)\) is recalculated, and the resulting density determines \(k_{\mathrm{on}}(x)\) for the next iteration.

We model CIL by assuming that cell--cell contact locally inhibits Rho GTPase activity by increasing the unbinding rate $k_{\mathrm{off}}$ during the cell--cell contact (Fig. \ref{fig:simulation_scheme}), which has duration $T_{\mathrm{CIL}}$. During contact we set %
\begin{equation}
k_{\mathrm{off}}(x,t)= k_{\mathrm{off}}\,[1+c(x)],
\qquad t_0 < t < t_0 + T_{\mathrm{CIL}},
\label{eq:koff(x, t)}
\end{equation}
where $c(x)$ is a function that indicates the local degree of cell--cell contact (Fig.~\ref{fig:free_vs_CIL}b). {Its functional form, illustrated in Fig.~\ref{fig:sketch_I_vs_c}, is}  %
\begin{align}
    c(x) &= c_0 \, S\!\left(x - (x_0 - \sigma)\right) \,
                   S\!\left((x_0 + \sigma) - x\right), \label{eq:c_x} \\
    S(u) &= \tfrac{1}{2}\left[1 + \tanh\!\left(\tfrac{u}{\omega}\right)\right], 
    \label{eq:S_x}
\end{align}
where $c_0$ is the inhibition strength, $\sigma$ the half-width of the contact zone,  
$x_0$ its center, and $\omega$ the boundary sharpness. This form is primarily just a convenient way to smoothly vary the local effect of CIL but can also be thought of as the accessible concentration of a signal on the contacting cell~\cite{Wang2024}.
We assume a localized inhibition motivated by experimental observations of suppressed protrusion at cell–cell interfaces~\cite{CarmonaFontaine2008}; alternative formulations incorporating diffusively spreading inhibition are discussed in Appendix~\ref{appendix:CIL_inhibitor_diffusion}. It also is possible to induce repolarization of a cell by altering other rates locally, and we expect quantitative differences to appear if other choices were made \cite{Buttenschn2022}.

To reduce trial-to-trial variability, we ensure that the polarized peak forms at roughly the same membrane location in each run by imposing a transient local increase in activation  \(k_{\mathrm{on}}(x)\) from \(t_s^{(i)} = 30\) to \(t_f^{(i)} = 90\,\mathrm{s}\) (Fig. \ref{fig:free_vs_CIL}a):
\begin{equation}
k_{\mathrm{on}}(x,t)= k_{\mathrm{on}}(x) \,[1+c^{(i)}(x)],
\qquad t_s^{(i)} < t < t_f^{(i)},
\label{eq:koff_initial(x, t)}
\end{equation}
where $c^{(i)}(x)$ is the same smoothed step function (Eq.~\ref{eq:c_x}) but with parameters \(c_0^{(i)} = 0.1\), \(\sigma^{(i)} = 2\,\mu\mathrm{m}\), \(\omega^{(i)} = 0.5\,\mu\mathrm{m}\), centered at \(x_0^{(i)} = 10\,\mu\mathrm{m}\). 
This produces a single polarized peak at the center of the domain before contact, which relaxes to a steady state by \(t_0 = 100~\mathrm{s}\).

\begin{table*}[ht]
\centering
\caption{Simulation parameters. Unless otherwise stated, the  values listed here are used throughout the study.}
\label{Table:parameters}
\begin{tabular}{l p{6.5cm} c c}
\toprule
Parameter & Description & Dimension & Value \\
\midrule
\multicolumn{4}{l}{\textit{Kinetic parameters}} \\
\midrule
$k_0$ & Baseline GEF activity & $T^{-1}$ & $0.067~\mathrm{s}^{-1}$ \\
$k_1$ & Maximum Hill-function rate & $T^{-1}$ & $1~\mathrm{s}^{-1}$ \\
$k_{\mathrm{off}}$   & Unbinding rate & $T^{-1}$ & $1~\mathrm{s}^{-1}$ \\
$\tilde{K}$ & Saturation parameter & $L^{-1}$ & $1~\mu\mathrm{m}^{-1}$ \\
\midrule
\multicolumn{4}{l}{\textit{Diffusion parameters}} \\
\midrule
$D_{\mathrm{m}}$ & Membrane diffusion coefficient & $L^{2}T^{-1}$ & $0.1~\mu\mathrm{m}^{2}\,\mathrm{s}^{-1}$ \\
$D_{\mathrm{c}}$ & Cytosolic diffusion coefficient & $L^{2}T^{-1}$ & $10~\mu\mathrm{m}^{2}\,\mathrm{s}^{-1}$ \\
\midrule
\multicolumn{4}{l}{\textit{CIL parameters}} \\
\midrule
$\delta x$ & Contact asymmetry relative to polarity peak & $L$ & Varying \\
$\sigma$ & Contact half-width & $L$ & Varying \\
$T_{\mathrm{CIL}}$ & Contact duration & $T$ & Varying \\
$c_0$ & Contact strength & --- & Varying \\
$\omega$ & Contact smoothness & $L$ & $1~\mu\mathrm{m}$ \\
\midrule
\multicolumn{4}{l}{\textit{Simulation setup}} \\
\midrule
$\tilde{\rho}_{\mathrm{m}}^{\mathrm{(i)}}$ & Initial membrane Rho GTPase concentration  & $L^{-1}$ & $0.3~\mu\mathrm{m}^{-1}$~\cite{Mori2008} \\
$\tilde{\rho}_{\mathrm{c}}^{\mathrm{(i)}}$ & Initial cytosolic Rho GTPase concentration & $L^{-1}$ & $2~\mu\mathrm{m}^{-1}$~\cite{Mori2008} \\
$C$ & Total mean Rho GTPase concentration & $L^{-1}$ & $2.3~\mu\mathrm{m}^{-1}$~\cite{Mori2008} \\
$L$ & Membrane length & $L$ & $20~\mu\mathrm{m}$ \\
$\Delta t$ & Time step & $T$ & $0.01~\mathrm{s}$ \\
$n_{\mathrm{bin}}$ & Number of spatial bins & --- & $100$ \\
\bottomrule
\end{tabular}
\end{table*}

\subsection{Domain discretization, parameter scaling with $N$ and comparison with the deterministic model}
\label{subsec: Domain Discretization and System Rescaling}

\begingroup\sloppy
In computing $k_\textrm{on}(x)$ (Eq.~\ref{eq:kon}), we 
 first determine the membrane-bound density $\rho_\textrm{m}(x) $
 from the positions $x_k$ of all active Rho GTPase proteins. To do so, we divide the membrane into $n_{\mathrm{bin}}$ bins  of width $\Delta x_{\mathrm{bin}}$. 
The local membrane-bound density $\rho_{\mathrm m}$
used in Eq.~\ref{eq:kon} is obtained as
\begin{equation}
  \rho_{\mathrm m}(j)=\frac{n_j}{\Delta x_{\mathrm{bin}}}, \qquad j =1,\dots,n_{\mathrm{bin}}, 
  \label{eq:rho_m}
\end{equation}
where $n_j$ denotes the number of membrane-bound proteins within bin $j$.

\endgroup

We want to vary the total number of proteins to tune the level of intrinsic noise (demographic noise)~\cite{vanKampen1992stochastic}. However, the deterministic wave-pinning model is highly sensitive to the mean Rho GTPase concentration~\cite{Mori2008, Camley2017-crawling, Camley2013}, so na\"{\i}vely changing $N$ holding all other parameters constant
can shift the system out of the polarization regime. In the earlier work~\cite{Walther2012}, this problem was addressed by simultaneously varying $N$ together with the cell width to keep the mean Rho GTPase concentration constant. 
{Our approach instead is to keep the cell geometry fixed while varying $N$, thereby tuning the magnitude of stochastic fluctuations. Although this also changes the concentration, we rescale the Hill saturation parameter $K_N$ to prevent the nonlinear feedback in Eq.~\ref{eq:kon} from trivially saturating and thus avoid shifting the system into or out of the polarized state.} 
We determine this scaling by requiring that, as $N\to \infty$, our stochastic model recovers the  deterministic wave-pinning equations, with
\begin{equation}
\begin{aligned}
        \frac{\partial}{\partial t}\tilde{\rho}_\mathrm{m}(x,t) 
        &= D_\mathrm{m}\frac{\partial^2}{\partial x^2}\tilde{\rho}_\mathrm{m}(x,t) 
        + k_\mathrm{on}(x)\tilde{\rho}_\mathrm{c} 
        - k_\mathrm{off}(x)\tilde{\rho}_\mathrm{m}, \\
        \frac{\partial}{\partial t}\tilde{\rho}_\mathrm{c}(x,t) 
        &= D_\mathrm{c}\frac{\partial^2}{\partial x^2}\tilde{\rho}_\mathrm{c}(x,t) 
        - k_\mathrm{on}(x)\tilde{\rho}_\mathrm{c} 
        + k_\mathrm{off}(x)\tilde{\rho}_\mathrm{m},
\end{aligned}
\label{eq: deterministic model}
\end{equation}
where $\tilde{\rho}_{\mathrm{m}}$ and $\tilde{\rho}_{\mathrm{c}}$ are the deterministic membrane and cytosolic densities, respectively. 
These deterministic equations conserve  total protein number, so that the spatial integral of the total density remains constant in time:
\begin{equation}
    \int_0^L [\tilde{\rho}_\textrm{m}(x,t) +  \tilde{\rho}_\textrm{c}(x,t)] \,dx \equiv  C L,
    \label{eq:integrated concentration LC}
\end{equation}
where $C$ is the mean total Rho GTPase concentration on the 1D domain of length $L$. 

In the stochastic model, $\rho_{\text{m}}$ and $\rho_{\textrm{c}}$ denote densities (number of proteins per length), defined as in Eq.~\ref{eq:rho_m}. Accordingly, the total copy number is
\begin{equation}
    \int_0^L [\rho_\textrm{m}(x,t) +  \rho_\textrm{c}(x,t)] \,dx = N.
\end{equation}
For consistency between the deterministic and stochastic descriptions, it is necessary that
$ \rho_{\mathrm m}=\alpha_N \tilde{\rho}_{\mathrm m},\, $
$\rho_{\mathrm{c }}=\alpha_N \tilde{\rho}_{\mathrm{c }}$, 
with $\alpha_N$
\begin{equation}
    \alpha_{N} = \frac{N}{C \, L}.
    \label{eq:scaling}
\end{equation}
The Hill saturation constant must then be rescaled  by the same factor, so that
\begin{equation}
    K_N = \alpha_N \tilde{K}.
\end{equation}
{with $\tilde{K}$ a fixed value (Table~\ref{Table:parameters}).}

\section{Results}

We applied the stochastic simulation framework described in Sec.~\ref{sec:methods} 
to quantify how intrinsic noise from finite Rho GTPase copy numbers influences cell migration and CIL outcomes.  %
Our analysis addresses two central 
questions: 
\textbf{(i)} how Rho GTPase copy number governs the persistence of polarity in freely migrating cells (Sec.~\ref{subsec: Free motile cells}), 
and \textbf{(ii)} how molecular noise alters the sensitivity of cells to contact cues (Sec.~\ref{subsec: CIL interaction}).

In our model, active Rho GTPase drives membrane protrusion --- we should think of the protein as, e.g. Cdc42 or Rac1. Regions of high $\rho_m(x)$ therefore correspond to the cell front and determine the direction of movement. To quantify this directionality, we define the ``instantaneous polarity vector'' as
\begin{equation} \label{eq:polarity}
    \Pvec(t) = \sum_{k \in \textrm{mem}} \bigl(\cos \phi_k, \;\sin \phi_k \bigr),
    \qquad
    \phi_k = \frac{2\pi x_k}{L},
\end{equation}
where $x_k$ denotes the position of the $k$-th protein along the cell perimeter, and the sum runs over proteins in the membrane-bound state (\(k \in \textrm{mem}\)). %
This vector sum provides the simplest way to infer the cell's direction from a spatially distributed set of active proteins; this formulation naturally appears in studies of chemotaxis \cite{Hu2010,Hu2011} and galvanotaxis \cite{Nwogbaga2023,Nwogbaga2024,Nwogbaga2025}, reflecting the idea that the proteins locally drive protrusion \cite{Nwogbaga2023}.

We quantify the extent of polarity reorientation due to intrinsic noise or CIL via the signed angular difference $\Delta \theta(t)$ between 
the time-averaged polarity vectors at the time of contact $t_0$ and the instantaneous polarity vector at a later time $t$
\begin{equation}
    \tan \Delta\theta(t) = 
    \frac{\bar{\Pvec}(t_0) \times \Pvec(t)}
         {\bar{\Pvec}(t_0)\cdot\Pvec(t)}, \qquad t>t_0.
    \label{eq: delta theta}
\end{equation}
Here $\bar{\Pvec}(t_0)\equiv \frac{1}{5 }\int_{t_0-5}^{t_0}\Pvec(s)\,ds$ is the polarity
vector averaged over a window of five seconds, and the two-dimensional cross product is 
$\mathbf{u} \times \mathbf{v} \equiv u_x v_y - u_y v_x$.

In practice, $\Delta\theta (t)$ is computed using the two-argument arctangent function (\texttt{atan2}).%

\subsection{Free migrating cells: stochastic fluctuations and persistence}
\label{subsec: Free motile cells}
First, we simulate cells in the absence of any cell--cell contact. A transient increase in activation \(c^{(i)}(x)\) is applied to set the initial membrane location \(x_0^{(i)}\) of the polarity peak, as discussed in Sec.~\ref{Stochastic model of Rho GTPase dynamics}. In this setting, we examine how the stability and stochastic wandering of the polarity peak depend on the total copy number $N$.%

\subsubsection{Polarity peak stability as a function of copy number $N$}
\begin{figure}[ht]
  \centering
  \includegraphics[width=1.0\linewidth]{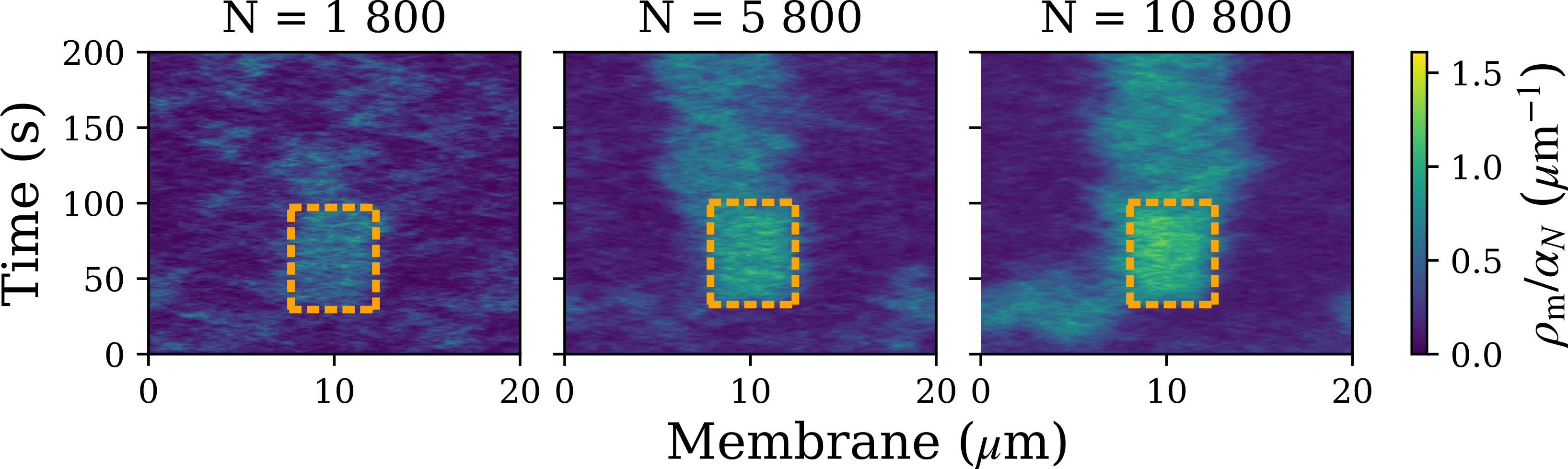}
    \caption{Kymographs of the normalized membrane concentration 
  $\rho_{\mathrm{m}}(t,x)/\alpha_{N}$ showing the spatiotemporal distribution 
  of membrane-bound proteins for three representative copy numbers $N$ in the absence of CIL.
  Yellow dashed lines indicate the time and spatial extent of the transient initial bias (Sec.~\ref{Stochastic model of Rho GTPase dynamics}).
}
  \label{fig:polarization for different N}
\end{figure}

The emergence and long-term persistence of a stable polarity to the cell depend strongly on the total copy number $N$ (Fig.~\ref{fig:polarization for different N}). At low copy number, $N \sim 2 \times 10^{3}$ molecules, no stable polarization peak persists
once the initial cue $c^{(i)}(x)$ is removed; instead, multiple transient peaks appear along the membrane. Increasing the copy number to  $N \sim 5 \times 10^{3}$ produces a detectable peak that exhibits noticeable wandering along the membrane. In contrast, for $N \gtrsim 10^{5}$,  a robust and nearly stationary polarization peak forms, and remains localized near the
position selected by the initial bias, with only small residual fluctuations. The stochastic wandering of the polarity peak as a function of copy number is
quantified below.

\subsubsection{Diffusion of the polarity peak at finite copy number}
\label{subsec: diffusion}

We expect that, in the absence of CIL, the persistence of motion in migrating cells is governed by the protein concentration dynamics shown in the kymographs in Fig. \ref{fig:polarization for different N}.
For sufficiently large $N$, the system forms a clearly polarized state --- a single peak of high protein concentration whose center of mass performs an unbiased random walk along the cell perimeter.
These fluctuations would then be captured by the diffusion coefficient of the polarization vector,
\(D_{\cal P}\)~\cite{Camley2014}.

{To compute \(D_{\cal P}\), we evaluate the mean-squared angular displacement averaged over \(N_{\mathrm{sim}} = 100\) independent realizations. Because \(\Delta\theta(t)\) is defined on \([-\pi,\pi)\) (Eq.~\ref{eq: delta theta}), we first unwrap each trajectory~\cite{harris2020array} to remove artificial jumps between \(-\pi\) and \(\pi\).}
The diffusion coefficient was obtained by fitting the linear regime of
\begin{equation}
    \langle \Delta \theta^{2}(t^{'}) \rangle = 2 D_{\cal P} t^{'},
\end{equation}
over the interval \(t' \in [0,\,100]~\mathrm{s}\).
Here \(t' = t - t_{0}\) is the time measured after the system reaches steady state at $t_0 =100 \, \textrm{s}$.

\begin{figure}[ht]
  \centering
  \includegraphics[width=1.0\linewidth]{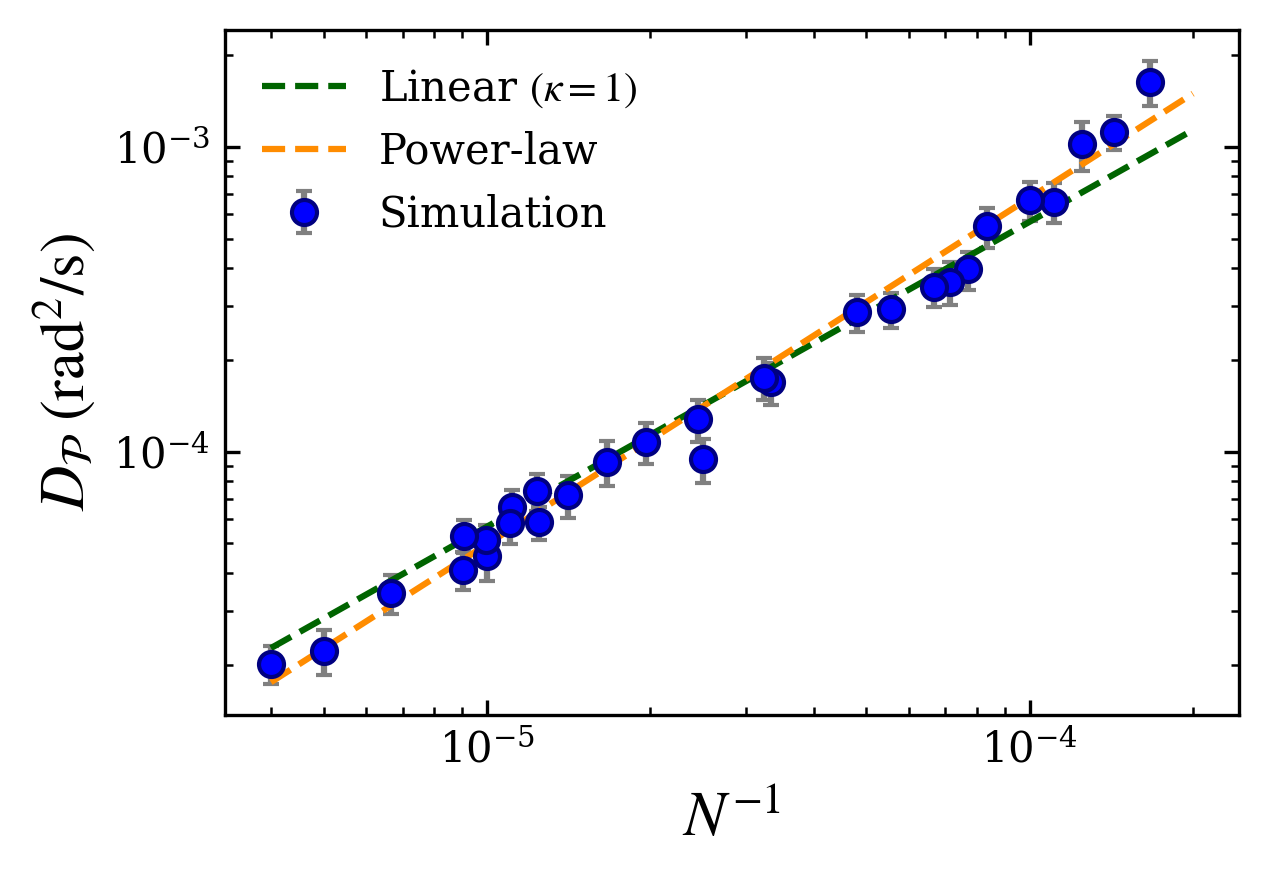}
    \caption{Log-log plot of the diffusion coefficient \(D_{\cal P}\) versus the inverse number of molecules \(N^{-1}\) for $N \in [5000, 25000]$. 
  The linear fit (green) is \(D_{\cal P} = a\,N^{-1}\), with \(a = 5.70 \pm 0.15\).
  The power-law fit (orange) is \(D_{\cal P} = b\,N^{-\kappa}\), with \(b = 3.16\pm 0.26  ,\ \kappa = 1.14 \pm 0.02\).  
  Standard errors are estimated by bootstrap resampling with $1000$ bootstraps.}
    \label{fig:DX_vs_N}
\end{figure}

We find that $D_{\cal P}$ decreases systematically with increasing copy number (Fig.~\ref{fig:DX_vs_N}). In the deterministic limit $N \to \infty$, the polarity axis would become stationary. At intermediate copy numbers, $N \sim (1.5\text{--}4)\times 10^4$, finite-$N$ fluctuations induce diffusion of the polarity peak with $D_{\cal P}$ decreasing from $\approx 3.8\times10^{-4} \, \mathrm{rad}^2/\mathrm{s}$ at the lower end of this range to $\approx 1.4\times10^{-4}\,\mathrm{rad}^2/\mathrm{s}$ at the higher end. The corresponding persistence time for directed motion, estimated as $\sim 1/D_{\cal P}$,  has a reasonable timescale for cell persistence of $0.7$--$2$ hours.

If the proteins were not interacting, we would predict that 
\(D_{\cal P} \sim N^{-1}\) (Appendix~\ref{appendix:1/N}).
Is the scaling of diffusion coefficient with $N$ strictly inverse linear or does it follow a more complicated dependence like a power law, e.g. 
$    D_{\cal P} \;\sim\; \,N^{-\kappa}$?
Fitting the data in Fig. \ref{fig:DX_vs_N} to a power law form yields $\kappa=1.14\pm0.02$, suggesting slightly stronger suppression of polarity fluctuations than the $N^{-1}$ scaling expected for independent particles. 
A value of $\kappa \simeq 1.14$ would be consistent with the idea that fluctuations in the polarity are suppressed due to positive feedback interactions. However, the deviation from $N^{-1}$ is modest in magnitude and the apparent differences between the two curves in Fig.~\ref{fig:DX_vs_N} are not large. Moreover, our model of thinking of the polarity as a single diffusing peak may not be appropriate at the lowest copy numbers, where the peak can split or change shape, effectively altering the strength of positive feedback~\cite{Walther2012}. Accordingly,  low-copy number regimes ($N \leq 10^4$ or so) do not necessarily follow the same scaling with $N$ as the asymptotic large-$N$ limit. 
If we fit only the range of $N>10^4$, we find $\kappa = 1.03 \pm 0.03$, which is consistent with simple linear scaling of $N^{-1}$ at large $N$ within uncertainty.

\subsection{CIL interactions: stochastic reorientation and sensitivity}
\label{subsec: CIL interaction}

\begin{figure}[ht]
  \centering
  \includegraphics[width=1.0\linewidth]{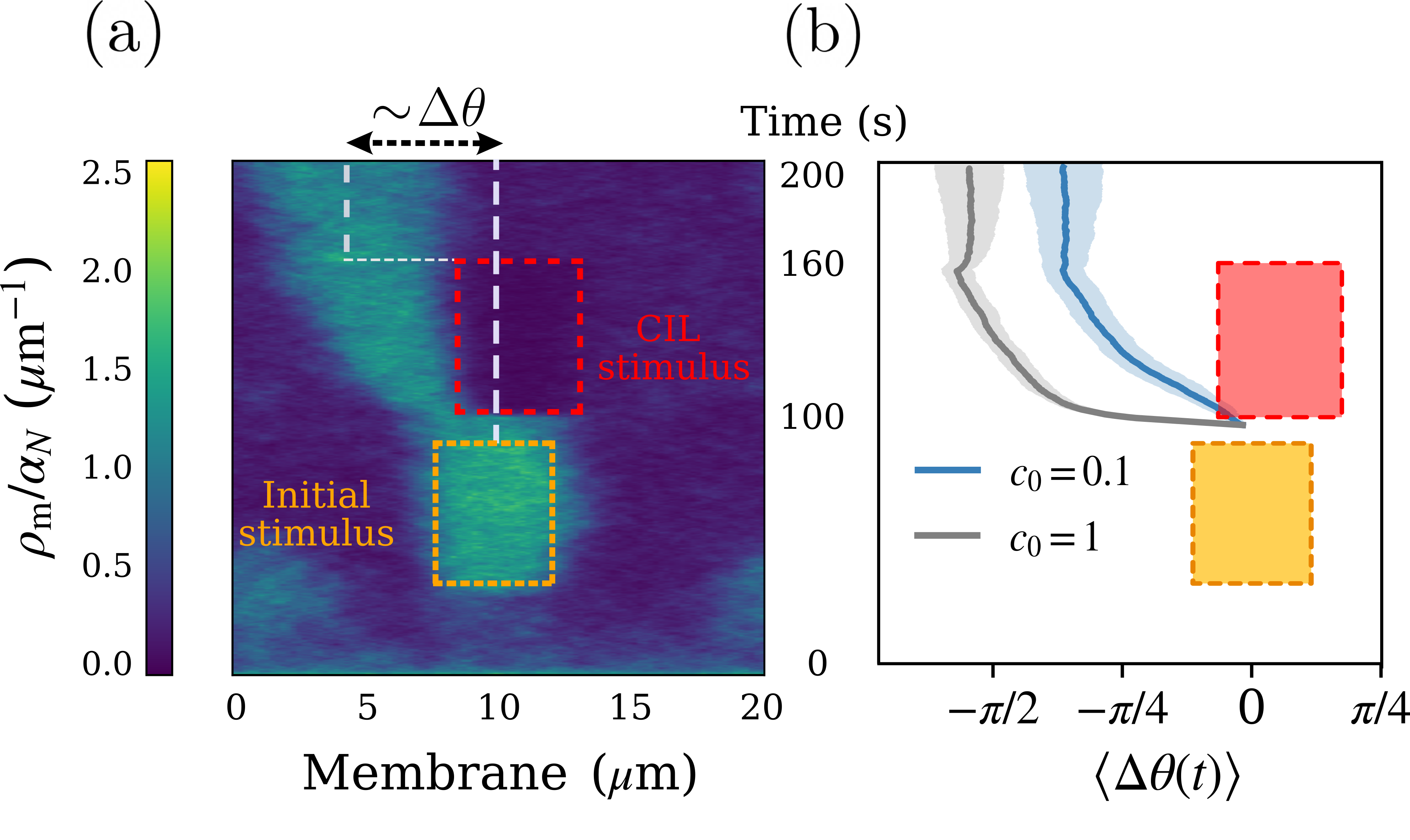}
  \caption{
  Illustration of a CIL event initiated at 
  $t_0 = 100\,\mathrm{s}$. Initial polarity prior to the CIL event is guided by a transient bias (yellow dashed lines; see Sec.~\ref{Stochastic model of Rho GTPase dynamics}).
  Parameters are $N = 15000$,
  $T_{\mathrm{CIL}} = 60\,\mathrm{s}$, 
  $\delta x = 1.5\,\mu\mathrm{m}$, $c_0 = 1$ and 
  $\sigma = 1.5\,\mu\mathrm{m}$. 
  \textbf{(a)} Kymograph of the normalized membrane concentration 
  $\rho_{\mathrm{m}}(t,x)/\alpha_{N}$ showing the redistribution of membrane-bound proteins (green intensity) during CIL stimulation. The angular displacement $\Delta \theta (t)$, computed from Eq.~\ref{eq:polarity},  can be related at a fixed time to the shift $\Delta x$ of the polarized peak along the membrane by $\Delta \theta = \frac{2\pi}{L}\Delta x$.
  \textbf{(b)} Time evolution of the mean angular shift 
  $\langle \Delta \theta (t) \rangle$ over $N_{\mathrm{sim}} = 100$
  for two different inhibition strengths with $c_0 = 0.1$ (blue line) and $c_0 = 1$ (red line). Shaded blue and red bands indicate the standard deviation of $\Delta\theta(t)$ across realizations. %
      }
  \label{fig:CIL_ex}
\end{figure}

Within our model, a cell responds to a CIL stimulus through a characteristic displacement of its polarized peak following contact (Fig.~\ref{fig:CIL_ex}). We quantify this behavior with the mean angular dispacement \(\langle \Delta\theta(t) \rangle\) of the polarity peak, computed from $N_{\mathrm{sim}} = 100$ stochastic simulations (Fig.~\ref{fig:CIL_ex}b). 

Cell--cell collisions are parameterized by four quantities (Fig.~\ref{fig:free_vs_CIL}b, Fig.~\ref{fig:sketch_I_vs_c}): 
\textbf{(i)} the contact asymmetry $\delta x = x_0 - x_1$, defined as the separation of the cell--cell contact point $x_0$ (see Eq.~\ref{eq:c_x}) from the polarity peak center $x_1$  
at the time of contact $t_0$;
\textbf{(ii)}  the half-width \(\sigma\) of the contact zone; \textbf{(iii)} the contact duration \(T_{\mathrm{CIL}}\); and \textbf{(iv)} the inhibition strength \(c_0\) (see Sec.~\ref{Stochastic model of Rho GTPase dynamics} for more details).   
The response is symmetric with respect to the contact location: CIL at \(\delta x\) and \(-\delta x\) elicit reorientations of equal magnitude and opposite sign.

To obtain a robust estimate of the reorientation angle for each realization \(i\), we compute a time-averaged angle over a window of duration
\(\Delta T = 10~\mathrm{s}\), after a \(t_{\mathrm{eq}} = 10~\mathrm{s}\)  equilibration period following the CIL signal:
\begin{equation}
    \overline{\Delta\theta}_i
    = \frac{1}{\Delta T}
      \int_{\tau}^{\tau + \Delta T}
      \Delta\theta_i(t)\, dt,
      \qquad \tau = t_{0} + T_{\mathrm{CIL}} + t_{\mathrm{eq}}.
    \label{eq: time average angle}
\end{equation}
The ensemble-averaged reorientation is then
\begin{equation}
    \langle \overline{\Delta\theta} \rangle
    = \frac{1}{N_\textrm{sim}}\sum_{i=1}^{N_\textrm{sim}} \overline{\Delta\theta}_i,
\label{eq: average N}
\end{equation}
and the corresponding standard deviation,
\begin{equation}
\delta_{\overline{\Delta\theta}}
    = \sqrt{\frac{1}{N_\textrm{sim}} \sum_{i=1}^{N_\textrm{sim}} \left( \overline{\Delta\theta}_i - \langle \overline{\Delta\theta} \rangle \right)^2}
    \label{eq: std dev N}
\end{equation}
quantifies the strength of stochastic fluctuations in the response.

Using this procedure, we can assess when the observed reorientation deviates from the deterministic prediction, and how finite copy-number fluctuations shape the cell’s sensitivity and migratory response to CIL.

\subsubsection{Distribution of cell reorientation angles in the stochastic
model}
\label{subsec:Distribution}

\begin{figure}[ht]
  \centering
  \begin{minipage}[t]{\linewidth}
    \centering
    \includegraphics[width=\linewidth]{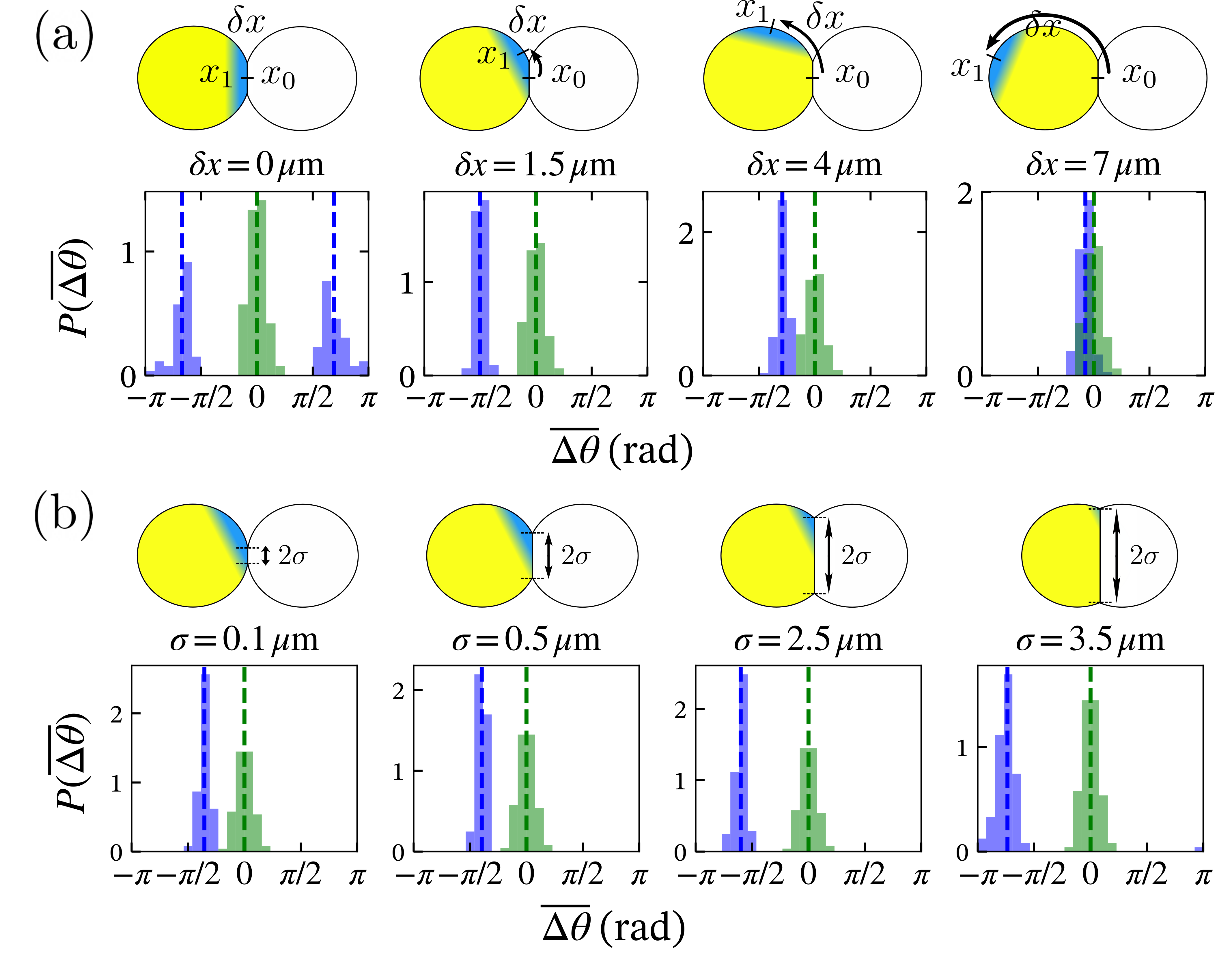}
  \end{minipage}
  \caption{Probability distributions \(P(\overline{\Delta\theta})\) for different contact
\textbf{(a)} asymmetries \(\delta x\) and
\textbf{(b)} half-widths \(\sigma\).
Green: control (no CIL); blue: CIL.
Distributions are obtained from
$N_{\mathrm{sim}} = 100$ stochastic simulations, with \(\overline{\Delta\theta}\) defined in
Eq.~\ref{eq: time average angle} and wrapped to $(-\pi, \pi]$.
Dashed lines indicate the corresponding mean reorientation angles
\(\langle \overline{\Delta\theta} \rangle\), defined in Eq.~\ref{eq: average N}.
Unless otherwise stated, fixed parameters are
\(T_{\mathrm{CIL}} = 60\,\mathrm{s}\), \(c_0 = 1\),
\(\delta x = 1.5\,\mu\mathrm{m}\),
\(\sigma = 1.5\,\mu\mathrm{m}\), and $N = 15000$.
  }
  \label{fig:Hist_asym_sigma}
\end{figure}

Given a cell--cell contact, how much does a cell reorient?  Unlike the deterministic model, which predicts a single rotation angle, stochastic simulations at finite \(N\) yield a distribution of reorientation angles \(P(\overline{\Delta\theta})\). 
We therefore quantify the CIL response by comparing the distributions \(P(\overline{\Delta\theta})\) 
 with and without CIL
across different contact parameters (Fig.~\ref{fig:Hist_asym_sigma}). 

For symmetric, head-on collisions ($\delta x = 0\,\mu\mathrm{m}$), the response splits into two symmetric peaks (Fig.~\ref{fig:Hist_asym_sigma}a), corresponding to clockwise and counterclockwise reorientation with equal probability. 
Increasing the contact asymmetry ($\delta x > 0\,\mu\mathrm{m}$) progressively suppresses one of the two peaks, leaving a single dominant orientation. As we show in the schematic in Fig.~\ref{fig:Hist_asym_sigma}a, larger asymmetries reduce the overlap between the contact region and the existing polarity peak, decreasing the separation between 
distributions in the presence and absence of CIL.
For sufficiently large $\delta x$, this separation vanishes, indicating loss
of CIL sensitivity  for highly asymmetric contacts. 
This reduced ability to distinguish CIL
when a cell is contacted at its ``back'' is expected within our relatively simple model, which incorporates only local inhibition of Rho GTPase activation at the site of cell--cell contact. Although the background concentration of active Rho GTPase at the back of the cell is nonzero, it is much lower than in the polarized peak, and its inhibition has little effect on cell polarity. More complicated models that incorporate a diffusing inhibitor created by cell--cell contact, as shown in Appendix \ref{appendix:CIL_inhibitor_diffusion}, can exhibit qualitatively different results. Similarly, we would also expect different results in this case if we modeled both Rac at the cell front and Rho at the cell back \cite{Buttenschn2022,Jilkine2011,Copos2020}.

The dependence on contact width $\sigma$ is more subtle
(Fig.~\ref{fig:Hist_asym_sigma}b). For narrow contacts 
($\sigma \lesssim 1.5\,\mu\mathrm{m}$), only a small portion of the polarized front is inhibited, producing a modest but reproducible rotation. Thus, the CIL distribution is shifted relative to the control, while its width remains small.
As $\sigma$ increases and the contact zone approaches the typical size of the polarity peak ($\sigma \sim 3.5\text{--}3.7\,\mu\mathrm{m}$), the inhibitory region covers most of the polarized peak. In this regime, small stochastic fluctuations in the position and shape of the peak translate into larger variability in the effective CIL signal.
Consequently, the CIL-induced distribution broadens, even though the mean reorientation remains well separated from the control.
The effects of contact strength and duration on the angular distribution are discussed in Appendix~\ref{appendix:CIL_strength_duration}.

\begin{figure}[ht]
  \centering
  \includegraphics[width=1\linewidth]{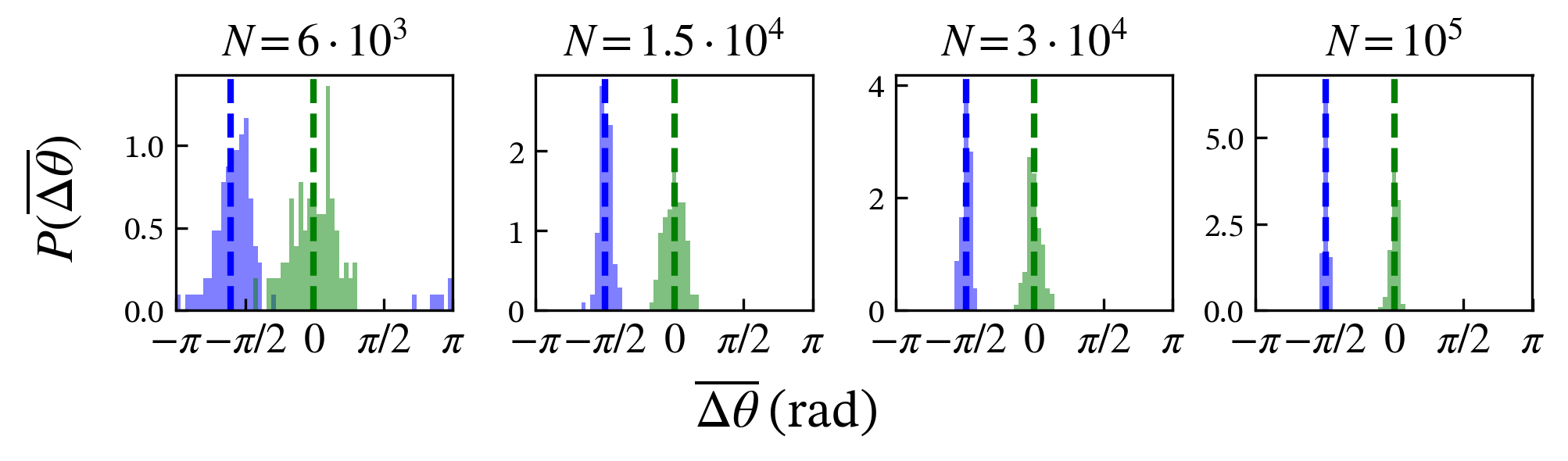}
    \caption{Probability distributions \(P(\overline{\Delta\theta})\) for different protein copy
numbers \(N\). Parameters are \(\sigma = 1.5\,\mu\mathrm{m}\),
\(c_{0} = 1\), \(T_{\mathrm{CIL}} = 60\,\mathrm{s}\), and
\(\delta x = 1.5\,\mu\mathrm{m}\).
}
    \label{fig:varied N}
\end{figure}

Protein copy number has the most dramatic effect on distribution width \(\delta_{\overline{\Delta\theta}}\).
At low copy number ($N = 6000$), the control and CIL distributions substantially overlap (Fig.~\ref{fig:varied N}) --- in this limit, some contact-induced reorientations are difficult to distinguish from spontaneous fluctuations. Conversely, increasing \(N\) narrows the distributions and drives the response toward the deterministic limit.

\subsubsection{Mean angular response across contact parameters}

\begin{figure}[ht]
    \centering
    \includegraphics[width=1.0\linewidth]{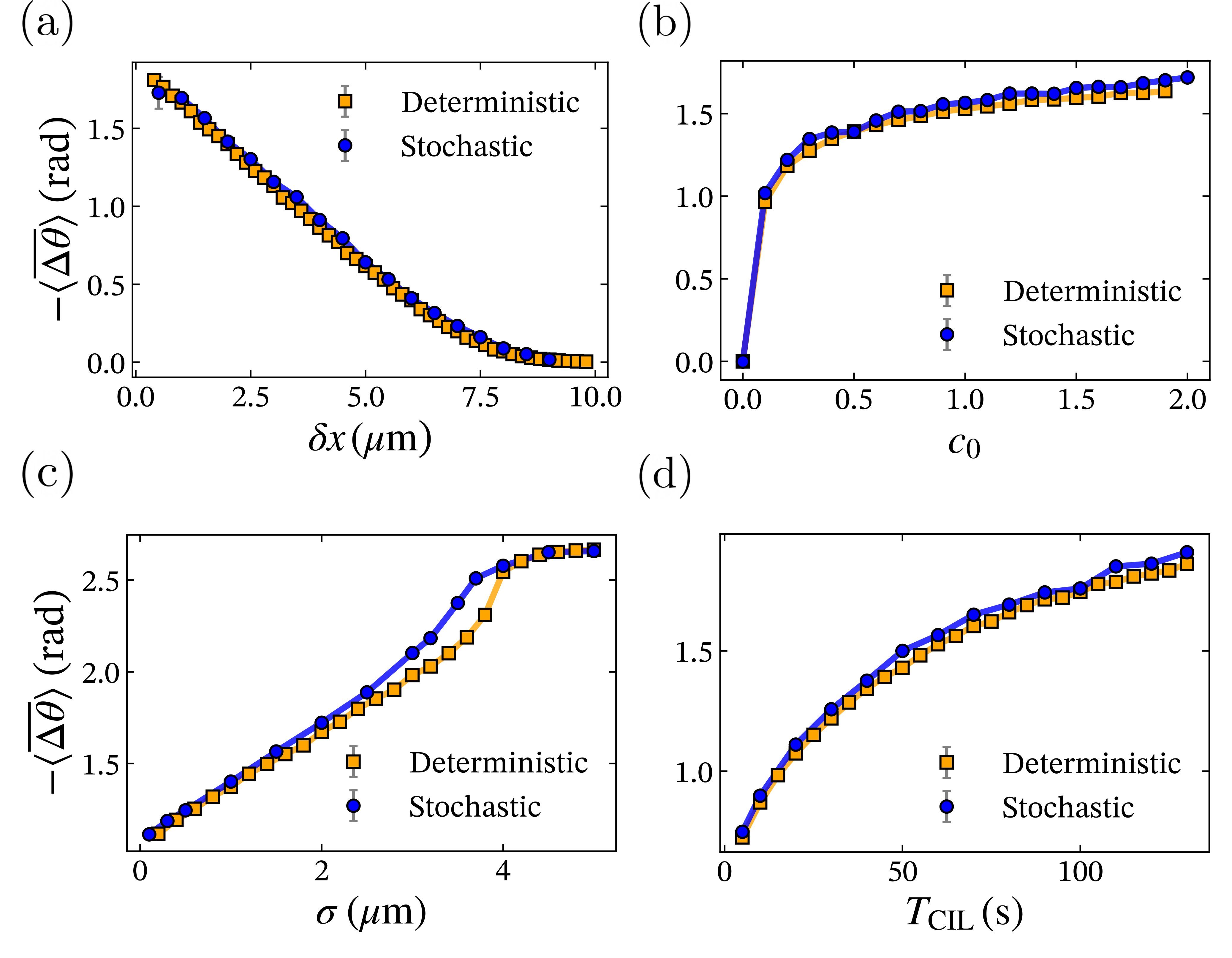}
    \caption{ 
    Angular shifts $\langle  \overline{\Delta\theta} \rangle$ generated by CIL with different \textbf{(a)} asymmetry \(\delta x\); \textbf{(b)} contact strength \(c_{0}\); 
\textbf{(c)} half-width \(\sigma\); \textbf{(d)} contact duration \(T_{\mathrm{CIL}}\).
Stochastic results are shown with blue circle markers (mean $\pm$ standard error of the mean, $\mathrm{SEM}=\mathrm{SD}/\sqrt{N_{\mathrm{sim}}}$). Deterministic points (orange squares) include the standard deviation across 50 runs with differing initial conditions --- here even the standard deviation is smaller than the marker size.
Unless otherwise stated, parameters are \(\sigma = 1.5\,\mu\mathrm{m}\), 
\(c_{0} = 1\), \(T_{\mathrm{CIL}} = 60\,\text{s}\), and \(\delta x = 1.5\,\mu\mathrm{m}\), \(N = 15000\). 
}
    \label{fig:Theta_all}
\end{figure}

How much, on average, does a cell--cell contact reorient a cell? How does this depend on the parameters controlling cell--cell contact, and is the average response simply given by the original deterministic wave-pinning model? 
We compute the ensemble-averaged reorientation angle $\langle \overline{\Delta\theta} \rangle$ (Eq.~\ref{eq: average N}) and compare it with the deterministic prediction obtained from simulating Eq.~\ref{eq: deterministic model}. These results are shown in Fig. \ref{fig:Theta_all}. {(We note that the deterministic results in  Fig.~\ref{fig:Theta_all} also have error bars from randomness in initial condition; these effects are much smaller than the variability in the stochastic model.})

We see in Fig.~\ref{fig:Theta_all} that the stochastic response is consistently slightly larger than the deterministic one, reflecting the influence of finite protein copy number and the resulting diffusive fluctuations of the polarity peak --- but this is a relatively weak effect and the average largely follows the deterministic result.

As the contact is displaced from the front toward the rear, changing asymmetry $\delta x$, the mean reorientation declines smoothly and becomes negligible once the inhibitory region no longer overlaps the polarized domain (Fig.~\ref{fig:Theta_all}a). Note that we exclude \(\delta x = 0 ~\mu\mathrm{m}\), where the mean reorientation \(\langle \overline{\Delta\theta} \rangle\) vanishes due to symmetry. Otherwise, a head-on collision could be misinterpreted as showing no effect of cell--cell contact, even though the distribution demonstrates a clear but symmetric response (Fig. \ref{fig:Hist_asym_sigma}a).

Varying the contact width $2\sigma$ produces a threshold-like trend (Fig.~\ref{fig:Theta_all}c). Broader inhibitory zones enhance reorientation up to a plateau at $\sigma \gtrsim 4\,\mu\mathrm{m}$, when the contact spans nearly half the membrane. Narrow contacts elicit weaker responses but remain detectable at $N=15000$ due to the relatively small variance at this copy number (Fig.~\ref{fig:varied N}).
At widths comparable to the intrinsic size of the polarity peak
($\sigma \approx 3$--$4\,\mu\mathrm{m}$), the stochastic response shows its largest deviation from the deterministic prediction. This same parameter regime is also where the distribution $P(\overline{\Delta\theta})$ becomes broad (Fig.~\ref{fig:Hist_asym_sigma}b), including events with angle changes approaching $\pi$. These large reorientation events may explain the discrepancy between the stochastic and deterministic models.

The inhibition strength $c_{0}$ has a saturating effect on reorientation --- increasing $c_{0}$ strongly increases the response but quickly reaches diminishing returns (Fig.~\ref{fig:Theta_all}b). In contrast, the dependence on contact time $T_{\mathrm{CIL}}$ is more gradual (Fig.~\ref{fig:Theta_all}d). %

\subsubsection{Cell sensitivity to CIL stimuli}
\label{subsec:Cell sensitivity across CIL stimuli}
How reliably can a cell distinguish a contact event from spontaneous
reorientation?
We quantify this \textit{contact sensitivity} using a  signal-to-noise ratio (SNR) as a heuristic separability metric,
\begin{equation}
\mathrm{SNR}
=\frac{\big(\langle \overline{\Delta\theta}\rangle_{\mathrm{free}}
-\langle \overline{\Delta\theta}\rangle_{\mathrm{CIL}}\big)^2}
{\delta_{\mathrm{free}}^{\,2}+\delta_{\mathrm{CIL}}^{\,2}}.
\label{eq:SNR}
\end{equation}
Here $\langle \overline{\Delta\theta} \rangle_{\mathrm{CIL}}$ and
$\langle \overline{\Delta\theta} \rangle_{\mathrm{free}}$ are the
ensemble-averaged reorientation angles with and without CIL,
respectively (Eq.~\ref{eq: average N}), and
$\delta_{\mathrm{CIL}}$ and $\delta_{\mathrm{free}}$ are the standard deviations (Eq.~\ref{eq: std dev N}).

\begin{figure}[t!]
  \centering
  \includegraphics[width=1.0\linewidth]{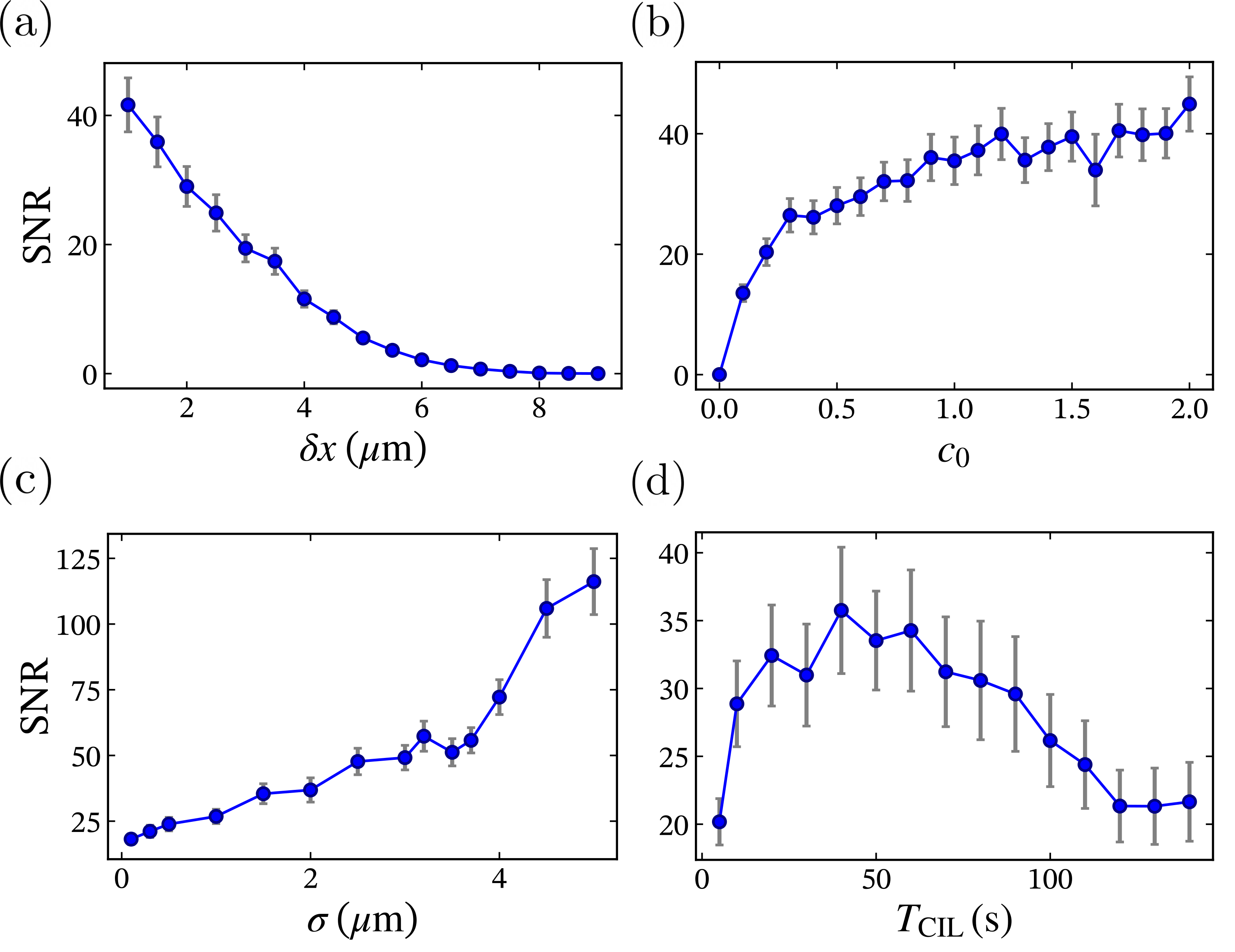}
    \caption{Signal-to-noise ratio (SNR) of the CIL response as a function of \textbf{(a)} asymmetry \(\delta x\); \textbf{(b)} contact strength \(c_{0}\); \textbf{(c)} half-width \(\sigma\); \textbf{(d)} contact duration \(T_{\mathrm{CIL}}\). 
    Unless otherwise specified, parameters are fixed at \(\sigma = 1.5\,\mu\text{m}\), \(c_{0}=1\), \(T_{\mathrm{CIL}} = 60\,\text{s}\), \(\delta x = 1.5\,\mu\text{m}\), and $N = 15000$. Error bars represent bootstrap uncertainties estimated from 1000 resamples.}
    \label{fig:SNR_all}
\end{figure}

Across the full parameter range,  $\mathrm{SNR} \gtrsim 1$, indicating that a single collision produces a reorientation that is typically larger than the intrinsic angular noise (Fig.~\ref{fig:SNR_all}). Thus, even at copy numbers of order $N\sim 10^{4}$, contact events are in principle detectable.
At higher, more physiological copy numbers ($N \sim 10^{5} - 10^{6}$), CIL
sensitivity is expected to be highly robust (Fig.~\ref{fig:CIL_N_convergence}b).

Contact parameters modulate this sensitivity in distinct ways
(Fig.~\ref{fig:SNR_all}). Increasing either the inhibition strength $c_{0}$ or the contact half-width $\sigma$ enhance the mean CIL-induced reorientation, leading to a monotonic increase in SNR (Fig.~\ref{fig:SNR_all}b, c).
Contact durations $T_{\mathrm{CIL}}$ initially increase sensitivity for short contacts ($T_{\mathrm{CIL}}\lesssim 60$~s), but
reduce it for longer interactions (Fig.~\ref{fig:SNR_all}d).  This
decrease arises because the variance of the no-CIL control  distribution
spreads at long times, while the post-CIL distribution remains the same (Fig.~\ref{fig:Hist_strength_duration}b), thereby reducing statistical separability.

In contrast, sensitivity decreases monotonically with increasing contact
asymmetry $\delta x$ (Fig.~\ref{fig:SNR_all}a). High SNR values at small \(\delta x \) indicate a robust, directed response when
the inhibitory contact overlaps the polarized front. Again, we are excluding from this plot the region of weak asymmetry (\(\delta x \lesssim 1 \, \mu\text{m}\)),  where the SNR is not a reliable measure due to a bimodal distribution of reorientation outcomes
(Fig.~\ref{fig:Hist_asym_sigma}a). For large \(\delta x\), where the contact occurs near the rear, the SNR approaches
zero, indicating a loss of directional CIL response.

\subsubsection{Convergence of the stochastic model to the deterministic limit (\texorpdfstring{$N \rightarrow \infty$} ))}

\begin{figure}[ht]
  \centering
  \includegraphics[width=1\linewidth]{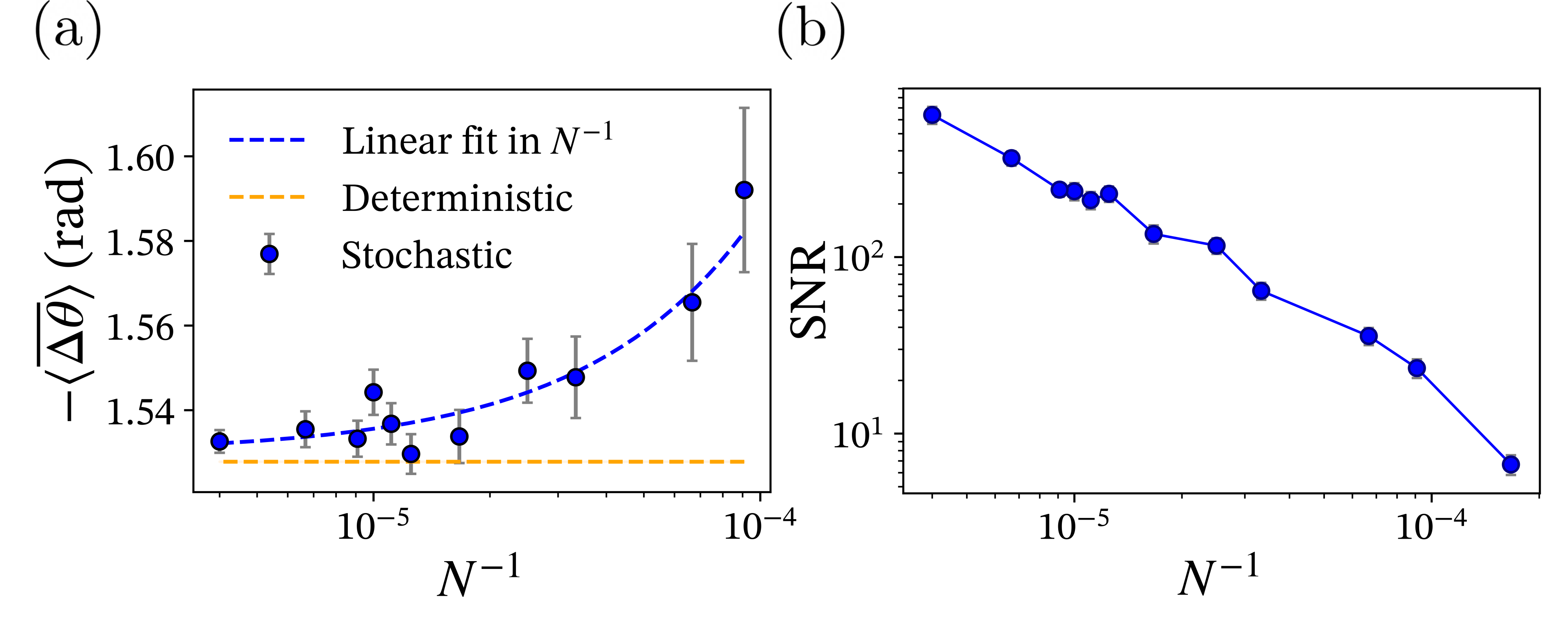}
  \caption{\textbf{(a)} Convergence of the stochastic mean angular shift
\(\langle \overline{\Delta\theta} \rangle\), defined in Eq.~\ref{eq: average N}, toward the deterministic limit as \(N^{-1} \to 0\).
The dashed blue line shows a linear fit of
\(\langle \overline{\Delta\theta} \rangle = a N^{-1} + b\) to the stochastic simulation data, obtained by weighted  least squares, {weighted with the SEM}. The error bars denote the SEM.
Extrapolation to \(N \to \infty\) yields
\(\langle \overline{\Delta\theta} \rangle = 1.530 \pm 0.005\,\mathrm{rad}\), 
in good agreement with the deterministic value \(\Delta\theta_{\mathrm{det}} = 1.528\,\mathrm{rad}\) (shaded orange band: \(\pm\) s.d. --- again, even the standard deviation of the deterministic model arising from noisy initial conditions is smaller than the stochastic uncertainty).
\textbf{(b)} Signal-to-noise ratio (SNR) of the CIL response as a function of protein 
copy number $N$. Fixed parameters: \(\sigma=1.5\,\mu\mathrm{m}\), \(c_0=1\), \(T_{\mathrm{CIL}}=60\,\mathrm{s}\), \(\delta x=1.5\,\mu\mathrm{m}\). Error bars are computed using bootstrapping with 1000 resamples.}
  \label{fig:CIL_N_convergence}
\end{figure}

We expect that the average CIL response of our stochastic model should approach the deterministic
prediction in the limit of a large number of proteins.
We test this by computing \(\langle\overline{\Delta\theta} \rangle\) 
for \(N = 10^{3}\)--\(10^{5}\) at fixed contact parameters.
As shown in Fig.~\ref{fig:CIL_N_convergence}, the mean angular reorentation converges smoothly toward the deterministic value; extrapolation to
\(N\rightarrow\infty\) yields an intercept in excellent agreement with the deterministic prediction.

Systematic deviations from the deterministic limit become apparent at low copy numbers ($N \sim 10^{3}$--$10^{4}$) (Fig.~\ref{fig:CIL_N_convergence}a). As we showed in Fig.~\ref{fig:varied N}, decreasing $N$ increases the width of the distributions of $P(\overline{\Delta\theta})$. This increase in variability increases the denominator of Eq.~\ref{eq:SNR}, decreasing SNR. Thus we see that molecular noise reduces the reliability of contact detection (Fig.~\ref{fig:CIL_N_convergence}b). %

\subsubsection{Acceleration vector analysis of CIL responses}

\begin{figure*}[ht]
  \centering
  \includegraphics[width=1\linewidth]{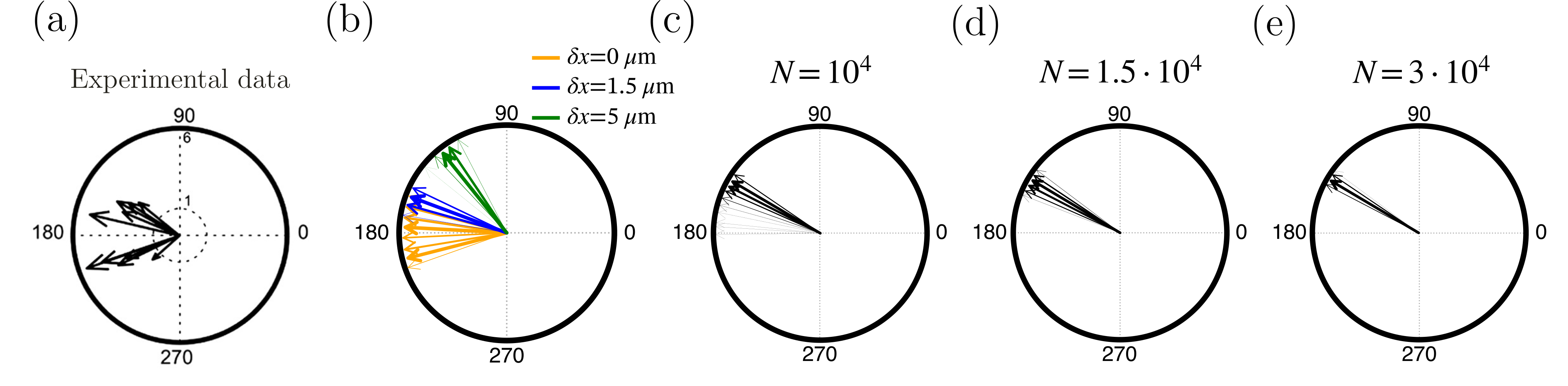}
  \caption{Comparison of acceleration vectors from simulations with experimental neural crest cell CIL~\cite{Theveneau2010}. 
  Angles are reported in a frame where the initial (pre-collision) cell movement is aligned along $0^\circ$ (to the right). %
\textbf{(a)} Experimental acceleration data. Arrows show the acceleration vector of an individual cell during a collision; both direction and magnitude are retained.
\textbf{(b-e)}
Arrows are normalized to unit length to show direction only; line width is proportional to the number of events in each angular bin.
\textbf{(b)} Simulations with varying CIL asymmetry. In our plotting, the contact orientation (defined by the CIL asymmetry) is at $0^\circ$ for $\delta x=0~\mu\mathrm{m}$, $27^\circ$ for $\delta x=1.5~\mu\mathrm{m}$, and $90^\circ$ for $\delta x=5~\mu\mathrm{m}$ 
(other parameters fixed: $\sigma=2.5~\mu\mathrm{m}$, $c_0=1$, $T_{\mathrm{CIL}}=80~\mathrm{s}$, $N=15000$; 70 angular bins).
\textbf{(c-e)}   Simulations with varied copy number, \(N = 10^4,\,1.5\times10^4,\,3\times10^4\) at fixed CIL parameters ($\sigma=2.5~\mu\mathrm{m}$, $c_0=1$, $T_{\mathrm{CIL}}=80~\mathrm{s}$, $\delta x=1.5~\mu\mathrm{m}$; 100 angular bins).
}
  \label{fig:comparison of CIL}
\end{figure*}

One of the standard tools used to quantify cell--cell collisions and infer the presence of contact inhibition in two-dimensional assays is to compute the acceleration of a cell in response to contact~\cite{Dunn1982, Theveneau2010, CarmonaFontaine2008}. Experiments show that in cell types that exhibit CIL, acceleration during collisions is on average biased away from the contact orientation, whereas in the absence of contact, or when CIL is inhibited, the acceleration distribution is more uniform, as expected from symmetry~\cite{Theveneau2010,CarmonaFontaine2008}. 
We plot typical experimental measurements in Fig.~\ref{fig:comparison of CIL}a as a polar plot, where $0^\circ$ denotes the pre-contact velocity orientation and $180^\circ$ corresponds to acceleration opposite the initial direction of motion. For approximately head-to-head collisions perfect CIL --- acceleration away from the contact --- will be in the  $180^\circ$ direction. 
Experimental CIL outcomes are highly variable from collision to collision, with a relatively broad spread of acceleration angles.

Does the directional spread of collision acceleration vectors correspond to the distribution of reorientation angles $\Delta \theta$ observed in our simulations? 
We compute the acceleration of a cell during a simulated CIL collision~\cite{Dunn1982} and show the results in Fig. \ref{fig:comparison of CIL}b-e.
Assuming that cells have a constant speed $v_0$, we write the velocity as \(\mathbf{v}(t)=v_0\,(\cos\Delta\theta(t),\,\sin\Delta\theta(t))\). With this parameterization, changing $v_0$ only rescales the magnitude of acceleration in a collision,  so we naturally focus on the orientation of acceleration vector.
In practice, we evaluate \(\mathbf{a}(t)\) using a central finite difference at the middle of the cell--cell contact time, 
which we call \(t_{*}\),
\begin{equation}
\mathbf{a}(t_*) \approx
\frac{\mathbf{v}(t_*+\Delta \tau)-\mathbf{v}(t_*-\Delta \tau)}{2\Delta \tau},
\end{equation}
where we use $\Delta \tau=\frac{T_{\mathrm{CIL}}}{2}$.  This estimate measures the average change in velocity over the interval
$t\in[t_*-\Delta\tau,\;t_*+\Delta\tau]$ %
and thus captures reorientation dynamics over an entire collision. We tune the parameters for Fig.~\ref{fig:comparison of CIL} slightly, changing the duration ($T_{\mathrm{CIL}}=80~\mathrm{s}$) and width of contact ($\sigma=2.5~\mu\mathrm{m}$), 
so that the scatter in accelerations is of the same order of magnitude as in the experimental data.
As we measure the polarity reorientation angle $\Delta\theta(t)$ with respect to the pre-contact direction (Eq.~\ref{eq: delta theta}), $\Delta\theta\simeq 0$ just before contact (Fig.~\ref{fig:CIL_ex}b).

In our simulations, head-on collisions ($\delta x = 0\, \mu\textrm{m}$) produce acceleration vectors biased directly away from the contact site, with a distribution similar to that observed experimentally  (Fig. \ref{fig:comparison of CIL}b). However, the acceleration response depends strongly on collision asymmetry: collisions with nonzero $\delta x$ generate a distribution of acceleration vectors that differ substantially  from the head-on case, consistent with the mean angular shifts observed in Fig.~\ref{fig:Theta_all}a.  Because the experimental data (Fig.~\ref{fig:comparison of CIL}a) combine collisions with many different $\delta x$ values, the observed spread of acceleration vectors can arise from a combination of stochastic variability at fixed $\delta x$ and heterogeneity arising from a different $\delta x$ across collisions (Fig~\ref{fig:Hist_asym_sigma}a).

Within our simulation, we can tune both $\delta x$ and the width of the distribution of acceleration vectors, which is controlled by $N$. At high copy number, the distribution of accelerations is effectively deterministic (Fig.~\ref{fig:comparison of CIL}e), but as the copy number is reduced, the distribution broadens substantially (Fig.~\ref{fig:comparison of CIL}c, d).  %

Our model suggests that both finite copy-number fluctuations and the variability of cell--cell contact (distribution of $\delta x$) can contribute to the heterogeneity of CIL responses, and together determine which source of noise dominates at different copy numbers $N$. This frustrates a direct comparison with the acceleration data commonly collected. A more quantitative test of our model --- or any detailed CIL model --- will require experiments that simultaneously quantify accelerations induced by cell--cell contact and the position of cell--cell contact relative to the cell front, since head-on collisions and glancing contacts will have different effects.

\section{Discussion}

We developed a stochastic model of cell polarity to
investigate how molecular noise and the properties of the cell--cell contact together control the outcome of CIL. We identify four key parameters: contact duration, contact location, contact width, and interaction strength (Fig.~\ref{fig:free_vs_CIL}b).
Our framework reveals two key regimes: one in which finite-copy-number noise dominates the response, and another in which variability is driven primarily by differences in contact geometry and strength.

\begin{figure}[ht]
  \centering
  \begin{minipage}[t]{\linewidth}
    \centering
    \includegraphics[width=1\linewidth]{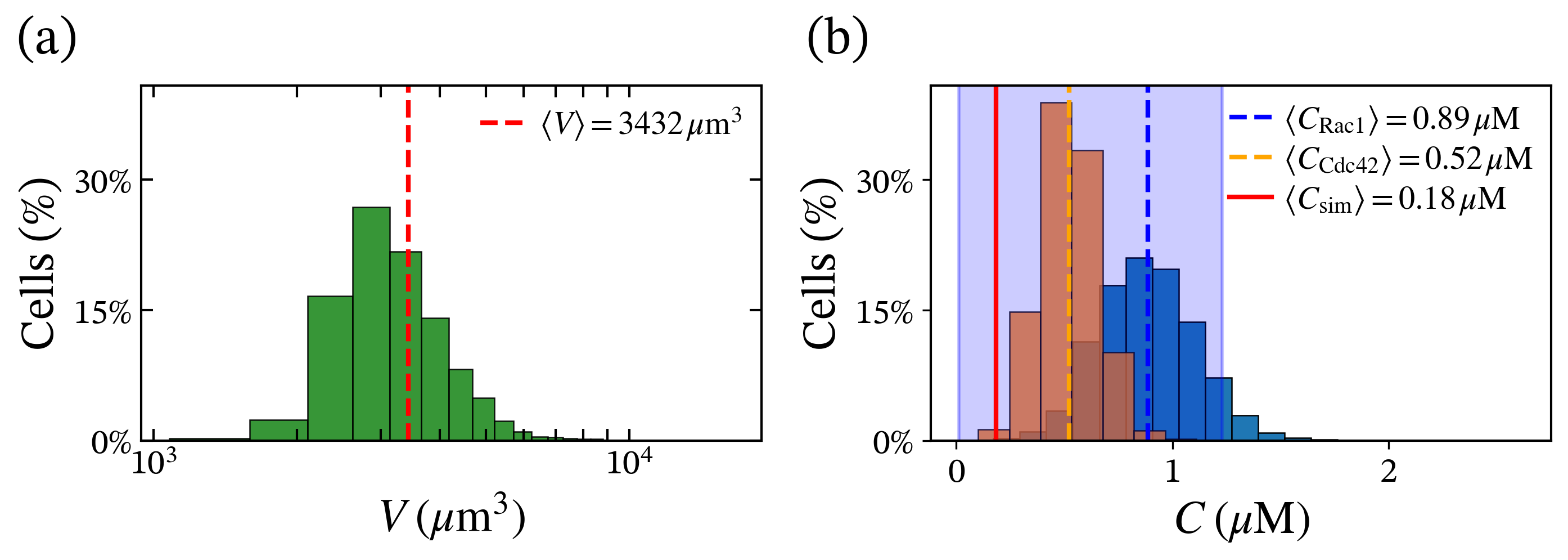}
  \end{minipage}
  \caption{Histogram of \textbf{(a)} S-HeLa cell volume~\cite{Kang2019}, with a mean of $3432~\mu\mathrm{m}^3$; and \textbf{(b)} the corresponding concentrations computed for Rac1 with $N=1.7\times10^6$ (blue) and Cdc42 with $N\sim 1\times10^6$ (orange), corresponding to $0.9~\mu\mathrm{M}$ and $0.5~\mu\mathrm{M}$, respectively. The blue shaded region indicates the concentration range explored in simulations, corresponding to $N\in[10^3,10^5]$. The red line marks the concentration corresponding to the most commonly used copy number, $N=1.5\times10^4$, which corresponds to $0.18~\mu\mathrm{M}$. }
  \label{fig:HeLa}
\end{figure}

Prior models of stochastic cell polarity have often assumed lower protein copy numbers (\( N\sim 10^2-10^3\)) to probe minimal requirements for polarity and symmetry breaking~\cite{Altschuler2008}. Here, we push toward more realistic values, though still within a very simplified model of Rho GTPase dynamics. {Quantitative measurements place small GTPase abundances in the millions per cell for mammalian systems; for example, Rac1 and Cdc42 in HeLa cells have been reported at approximately \(1.7 \times 10^{6}\) and \(1 \times 10^{6}\) molecules, respectively~\cite{Kulak2014}. Using an average HeLa cell volume of \(\sim 3400\,\mu\mathrm{m}^{3}\), together with the measured volume distribution shown in Fig.~\ref{fig:HeLa}a~\cite{Kang2019,Hatton2023}, these copy numbers correspond to concentration distributions shown in Fig.~\ref{fig:HeLa}b, with mean values of approximately \(0.9\,\mu\mathrm{M}\) for Rac1 and \(0.5\,\mu\mathrm{M}\) for Cdc42. At the same time, it is difficult to assign a single representative concentration even within one cell type, since protein abundance can vary substantially from cell to cell and across the cell cycle~\cite{Milo2013}.}  
In our simulations, {we explored total copy numbers for a single Rho-family GTPase in the range $N \in [10^3, 10^5]$  for a cell of radius \(R \approx 3\,\mu\mathrm{m}\) (volume \(\approx 135\,\mu\mathrm{m}^{3}\)). 
These values correspond to effective bulk concentrations of approximately \(0.012\text{--}1.23\,\mu\mathrm{M}\) (Fig.~\ref{fig:HeLa}b), reaching the correct order of magnitude for physiological Rho GTPase concentrations. Our default value for illustration in, e.g. Figs. \ref{fig:Theta_all}, \ref{fig:SNR_all}, is \(N = 1.5 \times 10^{4}\), corresponding to \(\approx 0.18\,\mu\mathrm{M}\), which lies near the lower end of the biologically relevant concentration range for Rho GTPases.}

At low copy numbers, we see that a minimum number of molecules is required to sustain a robust polarized peak (Fig.~\ref{fig:polarization for different N}), consistent with earlier work on the same model~\cite{Walther2012}. Even above this threshold, finite \(N\) noise causes slow diffusion of the polarization peak’s center of mass (Fig.~\ref{fig:DX_vs_N}), leading to reorientation over timescales set by the effective diffusion coefficient \(D_{\cal{P}}\).
Within the simulated range \(N \in [10^3, 10^5]\), we find \( D_{\cal{P}} \propto N^{-\kappa} \)  with \( \kappa \approx 1.14 > 1 \), 
lightly steeper than the \(N^{-1}\) scaling expected for non-interacting particles (Appendix~\ref{appendix:1/N}). This modest deviation may result from cooperative recruitment in the nonlinear feedback loop, similar to correlation effects observed in membrane domains~\cite{Sorkin2021}. {However, when we only fit to the data at large $N$, we find an exponent compatible with $\kappa = 1$ --- so the deviation from $\kappa = 1$ may reflect the dynamics at small $N$, where a single polarized peak can in rare cases split or take on a very different shape than in the larger-$N$ simulations. Our best simulations thus suggest that the asymptotic, large-$N$ exponent is compatible with $\kappa = 1$.}

Is the wandering of the Rho GTPase peak consistent with the degree to which a polarized eukaryotic cell loses its orientation over time? To assess whether intrinsic noise alone can drive spontaneous repolarization, we estimate the persistence time as \(\tau_{\cal{P}} \sim 1/D_{\cal{P}}\) (Fig.~\ref{fig:DX_vs_N})~\cite{Gail1970, Wu2014}. 
{For our typical value of $N = 1.5 \times 10^{4} $, this would correspond to a persistence time of $ 0.7$ hours. For $N = 2.6 \times 10^{4}$, the persistence time increases to a value close to the experimental estimate for fibroblasts, $t^{*} = 1.23$ h~\cite{Gail1970}. This suggests that intrinsic molecular noise in Rho GTPases alone may be sufficient to drive polarity loss on biologically relevant timescales.
} 
At the same time, additional sources of variability not included in our model are also likely to contribute, including fluctuations in upstream regulators (e.g., finite copy numbers and spatial dynamics of GEFs/GAPs), coupling to the cytoskeleton~\cite{Copos2020}, and stochastic receptor binding and unbinding. Consistent with this idea, recent work has shown that molecular-level fluctuations in models that account for the spatial dynamics of GEFs/GAPs can control the mobility of the polarization site in yeast~\cite{Ramirez_a_2021}.

Comparisons with experimental angular acceleration distributions in neural crest cells~\cite{Theveneau2010} (Fig.~\ref{fig:comparison of CIL}) also show that intrinsic noise may play a role. Experiments show acceleration vectors biased away from the site of contact, but with substantial angular spread~\cite{Dunn1982}. Direct comparison to our model is challenging because the experimental data pool collisions across all contact offset  $\delta x$, whereas our simulations assume a fixed $\delta x$ (colors in Fig.~\ref{fig:comparison of CIL}b). The resulting spread therefore arises from both the distribution of $\delta x$ in experiment and intrinsic stochastic fluctuations. 
{At the large copy numbers ($N \to 10^5$), where concentrations begin to approach the most likely biologically relevant values (Fig.~\ref{fig:HeLa}b), our results indicate that variation in $\delta x$, contact width or duration --- ``contact noise'' --- become the dominant factor, whereas intrinsic noise plays a secondary role.} 
Some recent time-lapse analyses of cell--cell collisions~\cite{Crossley2025, Prescott2021} have also focused on the relevance of the contact location, while others have focused on the role of relative velocities~\cite{Bruckner2021}.  Any future experimental tests of our approach or matching with our sort of detailed model will similarly require quantifying contact parameters (e.g., location, duration, and width of cell--cell contacts).
Understanding the origin of variability in CIL will also require a larger sample size than the typical tens of cells seen in Fig. \ref{fig:comparison of CIL}a. 

\begin{acknowledgments}
The authors acknowledge support from NIH Grant No.~R35GM142847.
This work was carried out at the Advanced Research Computing at Hopkins (ARCH) core facility, which is supported by the National Science Foundation (NSF) Grant No.~OAC1920103. We thank Emiliano Perez Ipi\~na and Daiyue Sun for a close reading of the draft and useful comments. 
\end{acknowledgments}

\bibliography{references}

\appendix
\counterwithin{figure}{section}
\counterwithin{table}{section}
\renewcommand\thefigure{\thesection\arabic{figure}}
\renewcommand\thetable{\thesection\arabic{table}}

\section{The choice of the time step for stochastic and deterministic simulations.}
\label{appendix:varied_dt}

In our stochastic model, \(\Delta t\) is required to satisfy two conditions: \textbf{(i)} the probability of a molecule binding and unbinding within the same time step is negligible, and \textbf{(ii)} diffusive displacements on the membrane, where motion is slowest, remain smaller than the spatial bin size. These conditions are
\begin{equation}
   r_{\mathrm{max}} \Delta t \ll 1, 
   \qquad 
   \sqrt{2D_m\Delta t} \ll \Delta x_{\mathrm{bin}},
\label{eq: 2conditions for SS}
\end{equation}
where \(r_{\mathrm{max}}\) is the largest reaction rate in the system and \(\Delta x_{\mathrm{bin}} = L/n_{\mathrm{bin}}\) is the bin width of a membrane of length \(L\). In practice, we set \(\Delta t\) such that the typical membrane diffusive step, \(\sqrt{2D_m\Delta t}\), is approximately an order of magnitude smaller than \(\Delta x_{\mathrm{bin}}\).
We note that cytosolic proteins can potentially hop more than one spatial bin within a single time step. Although this could, in principle, affect time-step convergence, near-uniform cytosolic mixing mitigates such effects. 

For computational efficiency, the time step \(\Delta t\) used in the stochastic
simulations does not strictly satisfy
\(2D_c\, \Delta t \ll \Delta x_{\mathrm{bin}}^2\). This is justified by the near-uniformity of the cytosolic concentration.
To verify that this choice does not affect the reorientation response,
we evaluated the mean angle shift
$\langle  \overline{\Delta\theta} \rangle$ (Eq.~\ref{eq: average N}) and the SNR 
over a broad range of integration time steps \(\Delta t\) (roughly two orders of magnitude).

\begin{figure}[ht]
  \centering
  \includegraphics[width=\linewidth]{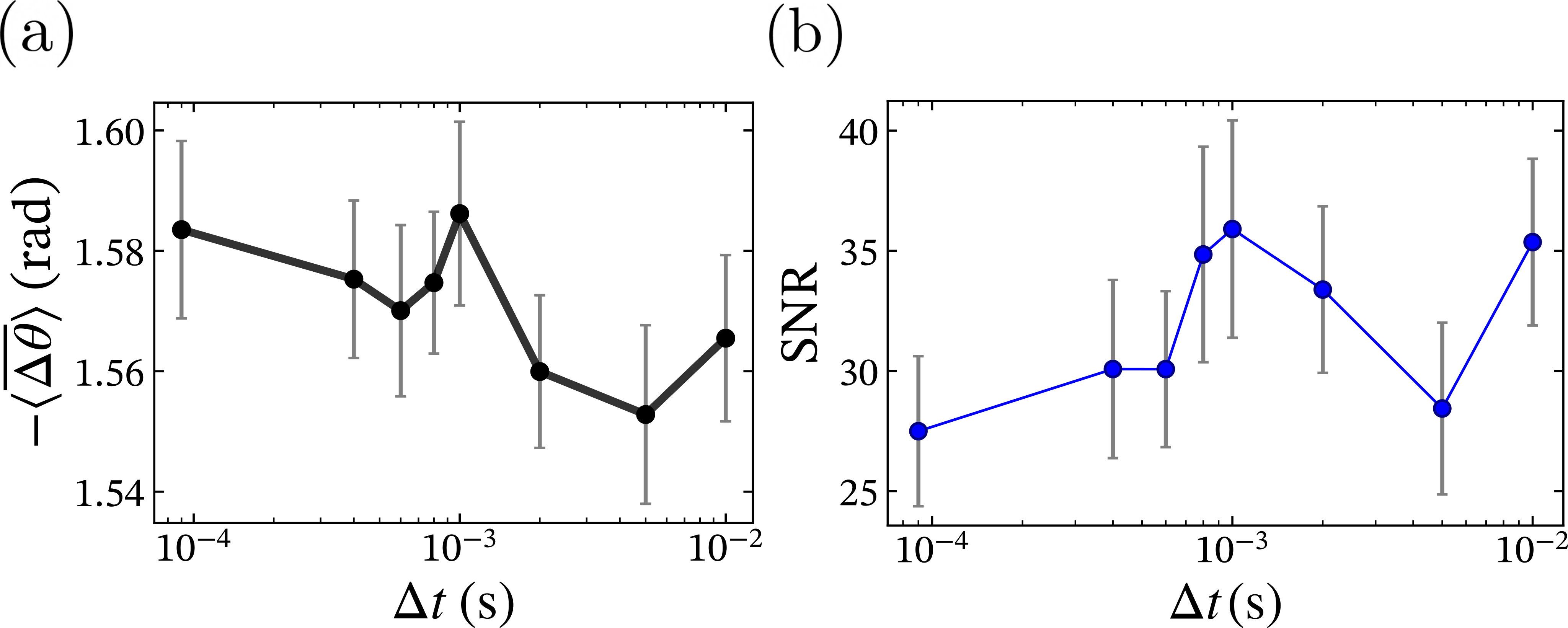}
  \caption{Robustness to the numerical integration time step \(\Delta t\).
  \textbf{(a)} Mean CIL-induced reorientation angle \(\langle \overline{\Delta\theta} \rangle\) (error bars: SEM).
  \textbf{(b)} Signal-to-noise ratio (SNR) of the CIL response. Error bars are estimated by bootstrap resampling with \(1000\) replicates.
  Fixed parameters: \(\sigma = 1.5\,\mu\mathrm{m}\), \(c_0 = 1\),
  \(T_{\mathrm{CIL}} = 60\,\mathrm{s}\), \(\delta x = 1.5\,\mu\mathrm{m}\), and \(N = 15000\). 
  }
  \label{fig:dt_varied}
\end{figure}

As shown in Fig.~\ref{fig:dt_varied}, the measured values exhibit no
systematic dependence on \(\Delta t\), and all variations remain within
statistical uncertainty. This indicates that the reported CIL response
is numerically robust to the choice of time discretization.

When we solve the deterministic model (Eq. \ref{eq: deterministic model}), we perform numerical integration on the same periodic spatial grid used in the stochastic simulations and apply a forward-time centered-space (FTCS) scheme with a time step of $\Delta t  \approx  1.7 \cdot 10^{-3} $ s. 
This time step is sufficient to ensure numerical stability and is approximately an order of magnitude smaller than that used in the stochastic simulations.

\section{\(1/N\) scaling of polarity diffusion for non-interacting particles}
\label{appendix:1/N}

Here we provide a brief reminder of why we might initially expect the scaling of the polarity diffusion coefficient
\(D_{\mathcal P}\) to be $1/N$ --- the limit of non-interacting membrane-bound particles.
Let \(\phi_k(t)\) denote the angular position of the \(k\)-th molecule diffusing along the membrane with angular diffusion coefficient \( D_\phi \). For a single Brownian particle,
\begin{equation}
    \langle [\phi_k(t)-\phi_k(0)]^2 \rangle = 2 D_\phi t .
\end{equation}

We define the angular position of the polarity peak as the center of mass of
the \(N_m\) membrane-bound molecules that form a polarized peak,
\begin{equation}
    \theta(t) = \frac{1}{N_m} \sum_{i=1}^{N_m} \phi_i(t) .
\end{equation}
We note that this linear average is not quite how we define the polarity direction in simulation (Eq. \ref{eq:polarity}) --- we expect these to be consistent if the particles are not too far apart. This is obviously not the case for noninteracting particles at long times. We do not present this as a serious model for what is happening in the full model but as a reminder of the unavoidable scaling with $N$ that occurs simply from averaging independent processes.

The effective diffusion coefficient \( D_{\mathcal{P}} \) of the polarization peak is obtained from  the
mean-squared displacement of \(\theta(t)\),
\begin{equation}
    \mathrm{MSD}_{\mathcal P}(t)
    = \left\langle [\theta(t)-\theta(0)]^2 \right\rangle .
\end{equation}

Assuming statistical independence of particle trajectories and setting
\(\theta(0)=0\), the cross terms vanish and we obtain
\begin{align}
    \mathrm{MSD}_{\mathcal{P}}(t) &= \left\langle \left( \frac{1}{N_{\mathrm{m}}} \sum_{i=1}^{N_{\mathrm{m}}} \phi_i(t) \right)^2 \right\rangle =\\
    &= \frac{1}{N_{\mathrm{m}}^2} \sum_{i=1}^{N_{\mathrm{m}}} \langle \phi_i^2(t) \rangle  = \frac{1}{N_{\mathrm{m}}^2}\, N_{\mathrm{m}} \, 2 D_\phi t  = 2 D_{\mathcal P} t
\end{align}

Thus, in the non-interacting limit,
\begin{equation}
    D_{\mathcal P} = \frac{D_\phi}{N_{\mathrm{m}}}.
\end{equation}

The typical number of membrane-bound molecules scales with total copy number, \(N_{\mathrm m} \sim N\), so we would expect with our full simulation, for noninteracting particles, we'd see $ D_{\mathcal P} \sim 1/N$.

\section{Distribution of angular reorientation in stochastic model}
\begingroup\sloppy

\label{appendix:CIL_strength_duration}

The strength of CIL strongly influences the mean angular reorientation
\(\langle \overline{\Delta\theta} \rangle\). Increasing the inhibition strength \(c_0\)
shifts the distribution \(P(\overline{\Delta\theta})\) toward larger
reorientation angles, while leaving its width essentially unchanged
(Fig.~\ref{fig:Hist_strength_duration}a).

Varying the duration of the CIL stimulus has a similar effect: longer
contact times increase the mean reorientation angle
(Fig.~\ref{fig:Hist_strength_duration}b), but the spread of the CIL
distribution again remains nearly constant. In contrast, the control
distribution (no CIL) broadens with time, reflecting diffusion of the
polarity peak in the absence of a stimulus.

\begin{figure}[t!]
  \centering
  \begin{minipage}[t]{\linewidth}
    \centering
    \includegraphics[width=\linewidth]{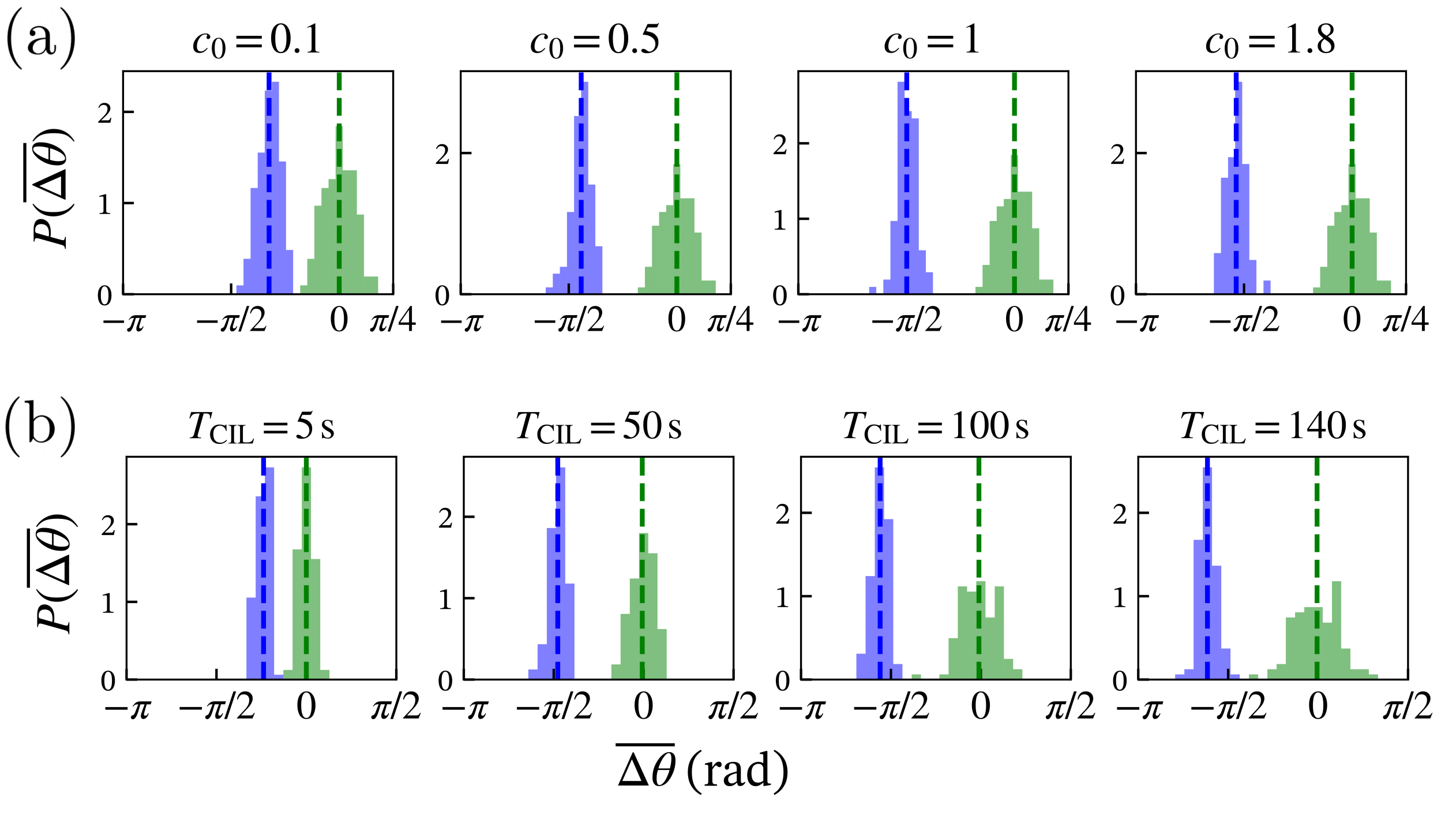}
  \end{minipage}
  \caption{Probability distributions \(P(\overline{\Delta\theta})\) for varied
  \textbf{(a)} contact strength \(c_0\) and 
  \textbf{(b)} contact duration \(T_{\mathrm{CIL}}\).
  Unless otherwise stated, fixed parameters are: \(T_{\mathrm{CIL}} = 60\,\mathrm{s}\), \(c_0 = 1\), \(\delta x = 1.5\,\mu\mathrm{m}\), 
  \(\sigma = 1.5\,\mu\mathrm{m}\), \(N = 15000\).}
  \label{fig:Hist_strength_duration}
\end{figure}

\section{Diffusive inhibition model for CIL}
\label{appendix:CIL_inhibitor_diffusion}

To examine how diffusion of an inhibitory signal shapes CIL, we consider an extension of our main model. Instead of assuming that a localized cell--cell contact of width $2\sigma$ directly inhibits Rho GTPase activation, 
we assume that it generates an inhibitory activity $I$ on the membrane, which can then spread beyond the initial contact region. We define \(I(x,t)\) as a scaled concentration of membrane components in an inhibitory state, which spreads by diffusion of inhibitor $D_I$ and decays with rate $\gamma$, e.g., 
$I$ is generated by contact-induced phosphorylation of a membrane species followed by dephosphorylation at some constant rate. 
The inhibitor profile $I(x)$ is obtained by solving a  one-dimensional reaction-diffusion equation on an unbounded domain \(x \in (-\infty, \infty)\):
\begin{equation}
  \left( \partial_{t} - D_I\partial_{x}^{2} + \gamma \right) I(t,x)
  = k \left[ \theta(x+\sigma_I) - \theta(x-\sigma_I) \right], \label{eq:Idiffusion}
\end{equation}
where \(\theta(x)\) is the Heaviside step function and \(\sigma_I\) denotes the half-width of the contact region.
The characteristic length scale over which inhibition spreads is \(\lambda = \sqrt{D_I/\gamma}\).

We adopt the following assumptions: \textbf{(i)} $\sigma_{I} \ll \lambda$, corresponding to narrow contacts with smooth boundaries; 
\textbf{(ii)}{$\gamma^{-1} \ll T_{I}$, so that $I$ relaxes to a steady-state profile much faster than the contact duration $T_I$} 
and 
\textbf{(iii)} \(\lambda \ll L\), where \(L\) is the cell perimeter, justifying neglect of finite-size effects and the use of an infinite domain.  
For representative parameters: \(D_I = 0.5\,\mu\text{m}^2/\text{s}\), \(\gamma = 0.1\,\text{s}^{-1}\), \(k = 0.15\,\text{s}^{-1}\) --- the resulting length scale is \(\lambda = 2.24 \,\mu\text{m}\).

We find the steady-state inhibitor profile
\begin{equation}
I(x) = \frac{k}{\gamma}
\begin{cases}
1 - e^{-\sigma_I/\lambda}\,\cosh\!\left(\dfrac{|x-x_I|}{\lambda}\right),
  & |x-x_I|<\sigma_I, \\[6pt]
\sinh\!\left(\dfrac{\sigma_I}{\lambda}\right)\,e^{-|x-x_I|/\lambda},
  & |x-x_I|>\sigma_I .
\end{cases}
\label{eq:I(x)}
\end{equation}
where we have also shifted the contact center from zero in Eq. \ref{eq:Idiffusion} to \(x = x_I\). 
 This solution replaces the localized step-like profile \(c(x)\) in Eq.~\ref{eq:koff(x, t)} with a smooth, exponentially decaying inhibitor field \(I(x)\) (Fig.~\ref{fig:sketch_I_vs_c}). When we compare with our results for the simple localized inhibition in the main paper, we choose the prefactor \(k/\gamma = c_0\). This sets the \textit{total inhibitory signal} (area under the profile): \(\int I(x)\,dx = \int c(x)\,dx\) to be matched as $\approx 2 c_0 \sigma$, at least in the limit $\sigma_I \ll \lambda$. 
The remaining parameters map directly to the model in the main text: the contact center is \(x_I\), the offset from the polarity peak at \(x_1\) is \(\delta x=x_I-x_1\), the contact half-width is \(\sigma_I\), and the signal duration is \(T_I\). This formulation allows us to examine how spatially extended inhibitory signaling modifies the CIL response.

\begin{figure}[ht]
  \centering
  \begin{minipage}[t]{\linewidth}
    \centering
    \includegraphics[width=0.8\linewidth]{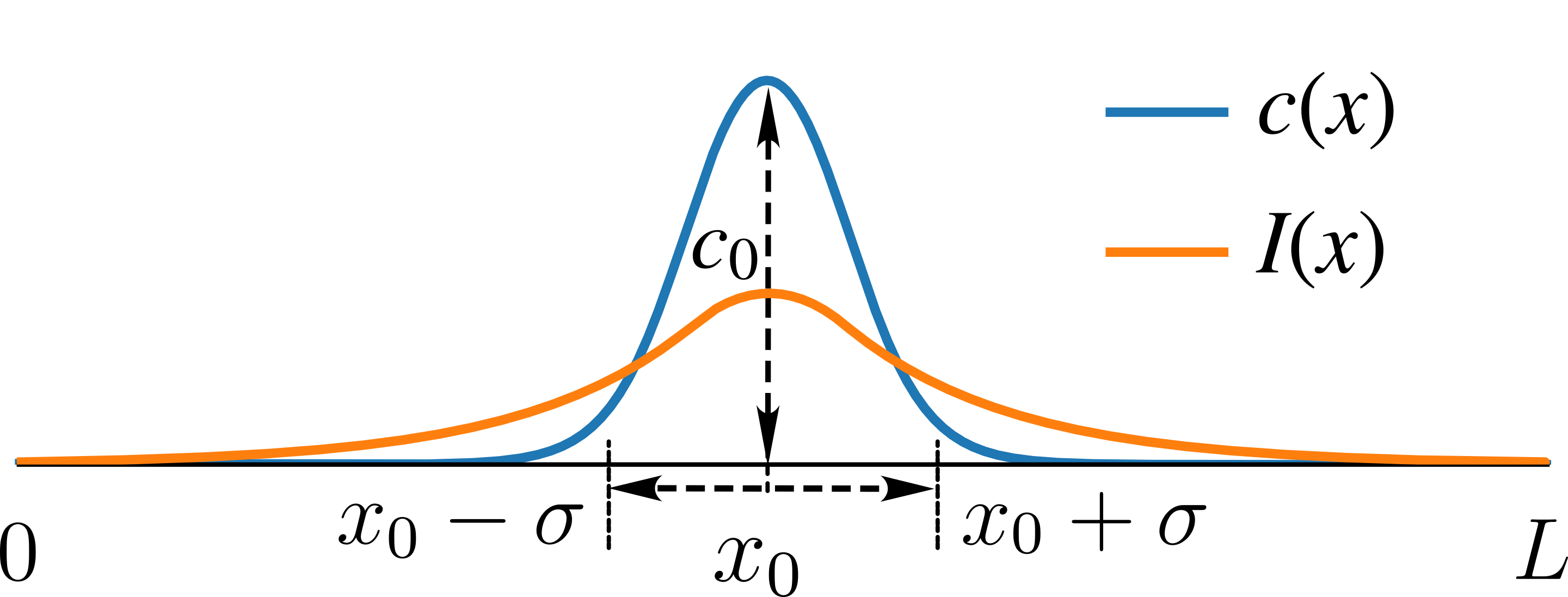}
  \end{minipage}
  \caption{Comparison between the localized cell--cell contact profile $c(x)$ (Eq.~\ref{eq:c_x}) and the spreading inhibition profile $I(x)$ (Eq.~\ref{eq:I(x)}). The parameters are $c_0 = k/\gamma = 1.5$, $x_0 = x_I = 10\,\mu\mathrm{m}$, $\sigma = \sigma_I
  = 1\,\mu\mathrm{m}$, and $\omega = 1$.}
  \label{fig:sketch_I_vs_c}
\end{figure}

For narrow contacts, both inhibition profiles $c(x)$ and $I(x)$ overlap with the polarized region and induce a reorientation of the polarity axis, but with quantitatively distinct responses (Fig.~\ref{fig:Sigma_disrt_dx1.5_C_I}). For the step-like profile $c(x)$ (blue), increasing the contact width over
the range $2\sigma = 0.2$--$1~\mu\mathrm{m}$ produces only a weak shift of the distribution, with the mean remaining close to $\langle \overline{\Delta\theta} \rangle \simeq -\pi/2$. In contrast, the model with \(I(x)\) shows a strong width dependence, with increasing
\(\sigma_I\) rapidly driving the response toward \(\langle \overline{\Delta\theta} \rangle \approx -\pi\), consistent with the \(\sim 180^\circ\) repolarization observed in some experiments~\cite{Labelle2025, Mayor2010}. We believe this \(\sim 180^\circ\) repolarization occurs in part because $I(x)$ is extended beyond the cell--cell contact. In this case, even once the cell's front has rotated away from the zone of cell--cell contact, there remains a graded inhibition --- the Rho GTPase is more inhibited toward the cell--cell contact than away from it. This graded effect will help drive the complete repolarization.
\begin{figure}[h!]
  \centering
  \includegraphics[width=1.0\linewidth]{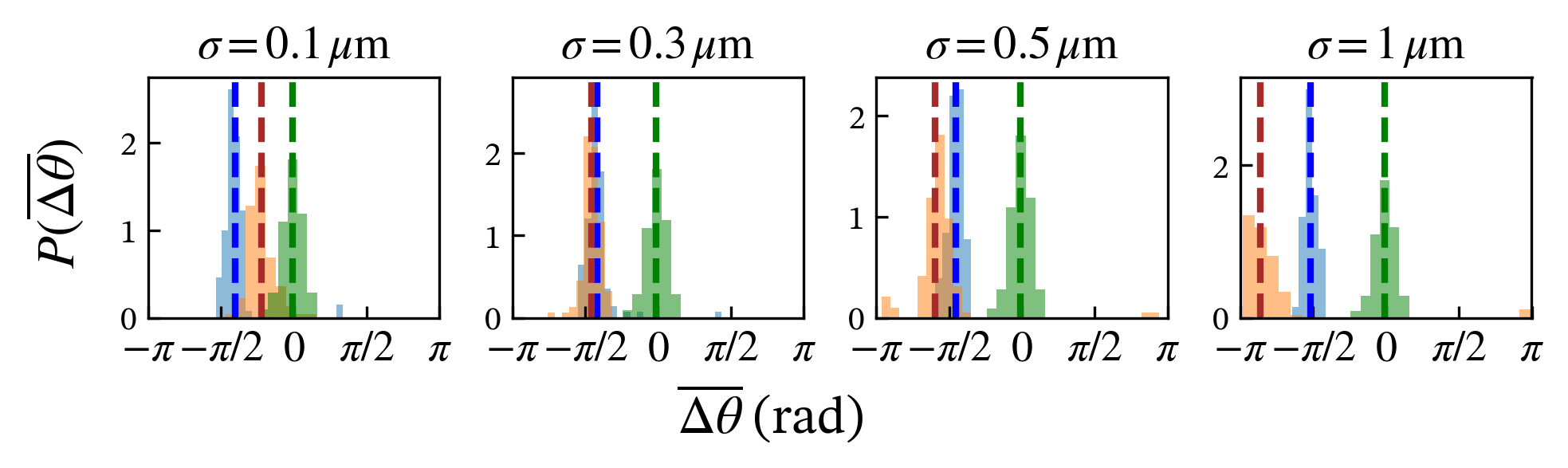}
    \caption{Probability distribution \(P(\overline{\Delta\theta})\) for different signaling profiles \(I(x)\) (orange) and \(c(x)\) (blue), compared to the control case without CIL (green). All simulations use a narrow contact region with width \(2\sigma \in (0.2, 2)\,\mu\mathrm{m}\), with $ \sigma = \sigma_{I}$. 
Fixed parameters: \(c_0 = \tfrac{k}{\gamma} =  1.5\), \(T_{\mathrm{CIL}} = T_{I} = 60~\mathrm{s}\), \(\delta x = \delta x_{I}  = 1~\mu\mathrm{m}\), and \(N = 15000\). 
}
   \label{fig:Sigma_disrt_dx1.5_C_I}
\end{figure}

Figure~\ref{fig:I_c} shows the effect of contact asymmetry for $2\sigma\simeq 1~\mu\mathrm{m}$.
The diffusive profile $I(x)$ induces a measurable mean
reorientation even for tail-to-head contacts (\(\delta x \approx 8\,\mu\text{m}\)), 
whereas the localized $c(x)$ produces essentially no response under the same conditions.  
The results in Figs.~\ref{fig:I_c} and \ref{fig:Sigma_disrt_dx1.5_C_I} show that diffusion of the inhibitor enhances the strength of CIL-driven reorientation and increases the spatial range over which contacts can effectively bias polarity, particularly for narrow or misaligned contacts.
\begin{figure}[t!]
  \centering
  \includegraphics[width=0.8\linewidth]{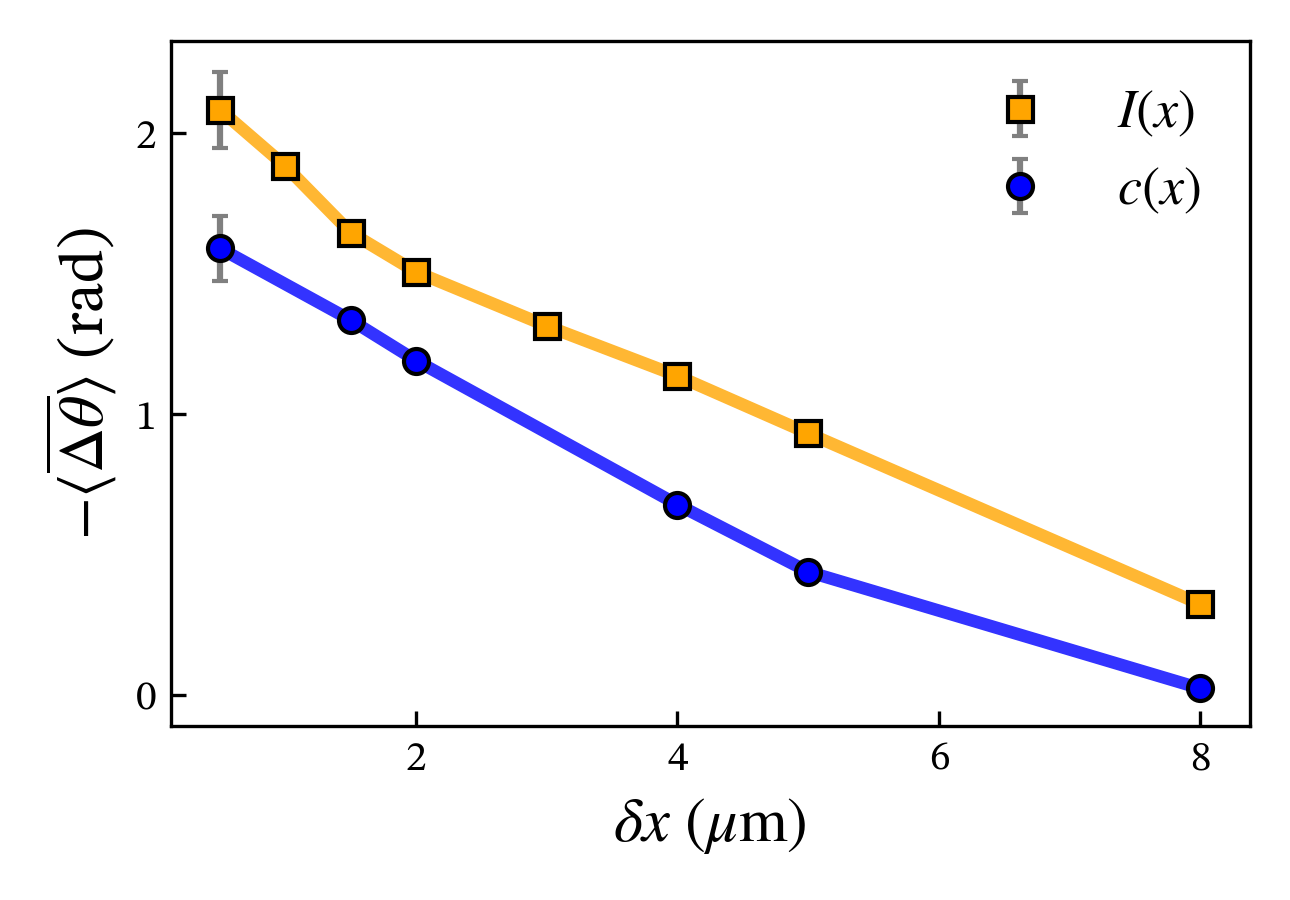}
  \caption{
    Mean reorientation angle as a function of contact asymmetry for different CIL signaling profiles. The orange curve corresponds to the diffusive inhibitor profile \(I(x)\), defined in Eq.~\ref{eq:I(x)}, while the blue curve shows the step-like profile \(c(x)\) defined in Eq.~\ref{eq:c_x}. 
    Fixed parameters:
    \(T_{\mathrm{CIL}} =  T_{I} = 60\,\text{s}\), 
    \(c_0 = \tfrac{k}{\gamma} =  1.5\), \(N = 15000\) 
    \(\sigma = \sigma_I = 0.5\,\mu\text{m}\).
  }
  \label{fig:I_c}
\end{figure}

\end{document}